\documentclass[%
 aip,
 amsmath,amssymb,
preprint,%
]{revtex4-1}

\usepackage{graphicx}
\usepackage{dcolumn}
\usepackage{bm}

\usepackage{microtype}
\usepackage[english]{babel}
\usepackage{lineno,hyperref}
\usepackage{txfonts}
\usepackage{microtype}
\modulolinenumbers[5]
\usepackage[ruled, lined, linesnumbered, commentsnumbered, longend]{algorithm2e}
\usepackage{comment}
\usepackage{footnote}
\usepackage{color}

\makeatletter
\def\@email#1#2{%
 \endgroup
 \patchcmd{\titleblock@produce}
  {\frontmatter@RRAPformat}
  {\frontmatter@RRAPformat{\produce@RRAP{*#1\href{mailto:#2}{#2}}}\frontmatter@RRAPformat}
  {}{}
}%
\makeatother
\begin{document}

\preprint{AIP/123-QED}

\title{Simulations of spin/polarization-resolved laser-plasma interactions in the nonlinear QED regime}

\author{Feng Wan}
\affiliation{Ministry of Education Key Laboratory for Nonequilibrium Synthesis and Modulation of Condensed Matter, Shaanxi Province Key Laboratory of Quantum Information and Quantum Optoelectronic Devices, School of Physics, Xi'an Jiaotong University, Xi'an 710049, China}

\author{Chong Lv}
\affiliation{Department of Nuclear Physics, China Institute of Atomic Energy, P. O. Box 275(7), Beijing 102413, China}%
 
\author{Kun Xue}
\affiliation{Ministry of Education Key Laboratory for Nonequilibrium Synthesis and Modulation of Condensed Matter, Shaanxi Province Key Laboratory of Quantum Information and Quantum Optoelectronic Devices, School of Physics, Xi'an Jiaotong University, Xi'an 710049, China}%

\author{Zhen-Ke Dou}
\affiliation{Ministry of Education Key Laboratory for Nonequilibrium Synthesis and Modulation of Condensed Matter, Shaanxi Province Key Laboratory of Quantum Information and Quantum Optoelectronic Devices, School of Physics, Xi'an Jiaotong University, Xi'an 710049, China}%

\author{Qian Zhao}
\affiliation{Ministry of Education Key Laboratory for Nonequilibrium Synthesis and Modulation of Condensed Matter, Shaanxi Province Key Laboratory of Quantum Information and Quantum Optoelectronic Devices, School of Physics, Xi'an Jiaotong University, Xi'an 710049, China}	

\author{Mamutjan Ababekri}
\affiliation{Ministry of Education Key Laboratory for Nonequilibrium Synthesis and Modulation of Condensed Matter, Shaanxi Province Key Laboratory of Quantum Information and Quantum Optoelectronic Devices, School of Physics, Xi'an Jiaotong University, Xi'an 710049, China}

\author{Wen-Qing Wei}
\affiliation{Ministry of Education Key Laboratory for Nonequilibrium Synthesis and Modulation of Condensed Matter, Shaanxi Province Key Laboratory of Quantum Information and Quantum Optoelectronic Devices, School of Physics, Xi'an Jiaotong University, Xi'an 710049, China}

\author{Zhong-Peng Li}
\affiliation{Ministry of Education Key Laboratory for Nonequilibrium Synthesis and Modulation of Condensed Matter, Shaanxi Province Key Laboratory of Quantum Information and Quantum Optoelectronic Devices, School of Physics, Xi'an Jiaotong University, Xi'an 710049, China}	

\author{Yong-Tao Zhao}
\affiliation{Ministry of Education Key Laboratory for Nonequilibrium Synthesis and Modulation of Condensed Matter, Shaanxi Province Key Laboratory of Quantum Information and Quantum Optoelectronic Devices, School of Physics, Xi'an Jiaotong University, Xi'an 710049, China}%

\author{Jian-Xing Li}\email{jianxing@xjtu.edu.cn}
\affiliation{Ministry of Education Key Laboratory for Nonequilibrium Synthesis and Modulation of Condensed Matter, Shaanxi Province Key Laboratory of Quantum Information and Quantum Optoelectronic Devices, School of Physics, Xi'an Jiaotong University, Xi'an 710049, China}	

\date{\today}

\begin{abstract}
	Strong-field quantum electrodynamics (SF-QED) plays a crucial role in ultraintense laser-matter interactions, and demands sophisticated techniques to understand the related physics with new degrees of freedom, including spin angular momentum. To investigate the impact of SF-QED processes, we have introduced spin/polarization-resolved nonlinear Compton scattering, nonlinear Breit-Wheeler and vacuum birefringence processes into our particle-in-cell (PIC) code. In this article, we will provide details of the implementation of these SF-QED modules and share known results that demonstrate exact agreement with existing single particle codes. By coupling normal PIC with spin/polarization-resolved SF-QED processes, we create a new theoretical platform to study strong field physics in currently running or planned petawatt or multi-petawatt laser facilities.
\end{abstract}

\maketitle

\section{Introduction} Laser-matter interactions can trigger strong-field quantum-electrodynamics (SF-QED) processes when the laser intensity $I_0$ reaches or exceeds $10^{22}~\mathrm{W/cm^2}$ \cite{Piazza_2012_Extremely,Bell2008}. 
For example, when the laser intensity is in the order of $10^{21}$-$10^{22}~\mathrm{W/cm^2}$, i.e., the normalized peak laser field strength parameter $a_0\equiv eE_0/m_ec\omega_0 \sim 10$, electrons can be accelerated to GeV energies \cite{Gonsalves_2019,Esarey_RMP} (with Lorentz factor $\gamma_e \sim 10^3$ or higher) in a centimeter-long gas plasma, where $-e,m_e$ are the charge and mass of the electron, respectively, $E_0, \omega_0$ are the field strength and angular frequency of the laser, respectively, and $c$ is the light speed in vacuum (here, for convenience, $\omega_{0} = \frac{2\pi c}{\lambda_0}$ and the wavelength of the laser $\lambda_0 = 1\mu \mathrm{m}$ are assumed). 
When the laser is reflected by a plasma mirror and collides with the accelerated electron bunch, the transverse electromagnetic (EM) field in the electron’s instantaneous frame can reach the order of $a' \simeq 2 \gamma a_0 \sim 10^4$-$10^5$. 
Such a field strength is close to the QED critical field strength (Schwinger critical field strength) $E_\mathrm{Sch} \equiv \frac{m_e^2c^3}{e\hbar}$, i.e., $a_\mathrm{Sch} = \frac{m_ec^2}{\hbar\omega_{0}} \simeq 4.1\times10^5$, within one or two orders of magnitude. 
In this regime, the probabilities of nonlinear QED processes are comparable to those of linear ones, and depend on three parameters as $W(\chi, f, g)$, with $\chi\equiv \frac{e\sqrt{(F_{\mu \nu}p^\mu)^2}}{m^3} \sim a'/a_\mathrm{Sch}$, $f\equiv \frac{e^2F_{\mu\nu}F^{\mu \nu}}{4m^4} \sim \frac{(\bm{a}_E^2 - \bm{a}_B^2)}{4a^2_\mathrm{Sch}}$ and $g \equiv \frac{e^2 F_{\mu \nu} F^{\mu \nu *}}{4m^4} \sim \frac{\bm{a}_E \cdot \bm{a}_B}{4a^2_\mathrm{Sch}}$ (here, $\bm{a}_{E,B}$ denote the normalized field strength of electric and magnetic component, respectively) \cite{Ritus_1985_Quantum,Baier1998}. 
For most cases of weak field ($a_0 \ll a_\mathrm{Sch}$) condition, $f,g \ll \chi^2$, and $W(\chi, f, g) \sim W(\chi)$, i.e., the probability only depends on a single parameter $\chi$. 
For electrons/positrons, the nonlinear Compton scattering (NCS, $e + n\omega_L \rightarrow e' + \omega_\gamma$) is the dominant nonlinear QED process in the strong field regime, and for photons, the nonlinear Breit-Wheeler pair production (NBW, $\omega_\gamma + n\omega_L \rightarrow e^+ + e^{-}$) is the dominant one, where $\omega_L, \omega_\gamma$ denote the laser photon and the emitted $\gamma$ photon, respectively, and $n$ is the photon absorption number.

Apart from these kinetic effects, the spin/polarization effects also arise with the possibility of generating polarized high-energy particle beams or when particles traverse large-scale intense transient fields in laser-plasma interactions. Classically, the spin of a charged particle will precess around the instantaneous magnetic field, i.e. $\mathrm{d}{\bf s}/\mathrm{d}t \propto {\bf B} \times {\bf s}$, where $\bf{s}$ denotes the classical spin vector \cite{Jackson_2021}. In storage rings, due to the radiation reaction, the spin of an electron/positron will flip to the direction parallel/antiparallel to the external magnetic field, i.e., the Sokolov-Ternov effect (an unpolarized electron beam will be polarized to a degree of $\sim 92.5\%$) \cite{sokolov-ternov}, and a similar process also occurs in the NCS \cite{sorbo_2017, Li2019, King2020}. Some recent studies have shown that with specific configurations, for example, when elliptically or linearly polarized lasers scatter with high-energy electron bunches (or plasmas), the polarization degree of the electrons can reach $90\%$ and be used to diagnose the transient fields in plasmas \cite{Li_2020_Polarized,Gong_2021}. Meanwhile, the photons created by NCS can be polarized, and when these polarized photons decay into electron/positron pairs, the contribution to the probability from polarization can reach $\sim 30\%$ \cite{WanPRR}, and will be inherited by the subsequent QED cascade. For example, in the laser-plasma/beam interactions, the polarization degree for linearly polarized (LP) photons is about $60\%$ or higher and for circular polarized (CP) $\gamma$ photons can reach 59\% when employing longitudinally polarized primaries\cite{Xue2020, King2020, Tang2020, Song_2022}.

Analytical solutions in ultraintense laser-matter interactions are scarce due to the high nonlinearity and complexity of the problem. Moreover, the micro-level processes such as ionization, recombination and Coulomb collisions, etc., coupled with the complicated configurations of lasers and plasmas make the explicit derivation almost impossible. Fortunately, computer simulation methods provide alternative and more robust tools to study those unsolvable processes even in more realistic situations \cite{Arber_2015}. In general, simulation methods for laser-plasma (ionized matter) interactions can be categorized as kinetic or fluid simulations, and specifically kinetic methods include the Fokker-Planck equation (F-P) (or the Vlasov equation for the collisionless case) and the particle-in-cell (PIC) method, and the fluid method mainly uses the magnetohydrodynamic equations (MHD) \cite{Birdsall_2018}. Among these methods, both F-P and MHD discretize the momentum space of particles and are prone to the nonphysical multi-stream instability, which may obscure the real physics, such as the emergence of turbulence, physical instabilities, etc. In comparison, the PIC method can provide much more detailed information on the discrete nature and intrinsic statistical fluctuations of the system, regardless of the stiffness of the problem. Therefore, the PIC method has been widely used in the simulation of ultraintense laser-plasma interactions \cite{Arber_2015, Birdsall_2018, Gonoskov_2015_Extended}.

Thanks to the emerging PIC simulation methods, the development of parallelism and large-scale cluster deployment, simulations of laser-plasma wakefield acceleration, laser-ion acceleration, THz radiation and also SF-QED, etc., have become accessible for general laser-plasma scientists \cite{OSIRIS, Burau_2010, Arber_2015, Derouillat_2018, Lapins}. However, the spin and polarization properties of the plasma particles and QED products are not widely introduced to the mainstream due to the lack of appropriate algorithms. In some recent studies, the spin and polarization resolved SF-QED processes have been investigated in the laser-beam colliding configurations and have shown that these processes are prominent in generating polarized beams \cite{WanPRR, King2020, Tang2020, Song_2022, Li2019, Li2022}. And these local-constant-approximated version of these probabilities can be readily introduced into any PIC code.

In this paper, we briefly review the common PIC simulation algorithms and present some recent implementations in spin/polarization averaged/summed QED. The formulas and algorithms for the spin/polarization-dependent SF-QED processes are given in detail and have been coded into our PIC code SLIPs (which stands for ``{\bf S}pin-resolved {\bf L}aser {\bf I}nteraction with {\bf P}lasma {\bf s}imulation code''). These formulas and algorithms presented in this paper, especially the polarized version, can be easily adopted by any other PIC code and used to simulate the ultraintense laser-matter interactions that are already available or will be achievable in the near future multi-petawatt (PW) to exawatt (EW) laser facilities \cite{Danson_2015_Petawatt}, such as Apollo \cite{Apollon, Zou_2015_Design}, ELI \cite{ELI}, SULF \cite{SULF} and SEL, etc.
Throughout the paper, Gaussian units will be adopted, and all quantities are normalized as follows: {time $t$ with $1/\omega$ (i.e., $t' \equiv t / (1/\omega) = \omega t$), position $x$ with $1/k = \frac{\lambda}{2\pi}$, momentum $p$ with $m_e c$, velocity $v$ with $c$,  energy $\varepsilon$ with $m_e c^2$, EM field $E, B$ with $\frac{m_e c \omega}{e}$, force $F$ with $m_e c \omega$, charge $q$ with $e$, charge density $\rho$ with $k^3 e$, current density $J$ to $k^3 e c$, where $\lambda$ and $\omega = \frac{2\pi c}{\lambda}$ are the reference wavelength and frequency, respectively.}

\section{PIC Algorithm}
\begin{figure}[ht]
\centering
\includegraphics[width=0.8\linewidth]{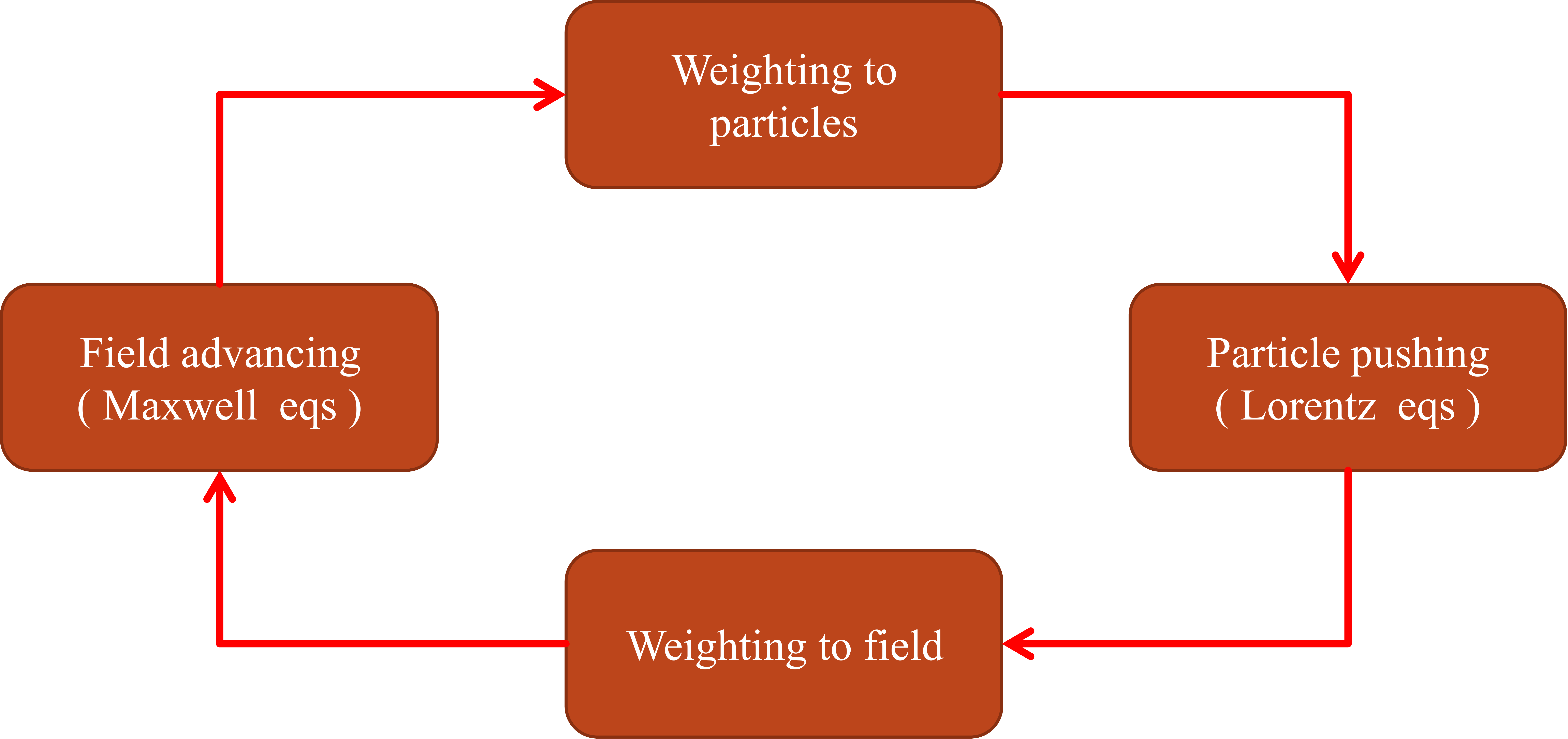}
\caption{Standard particle-in-cell (PIC) loop with four kernel parts.}\label{fig-loop}
\end{figure}

The simulation of laser-plasma interactions requires two essential components: the evolution of the EM field and the motion of particles. The corresponding governing equations are the Maxwell equations (either with $\bf{A}$-$\phi$ or $\bf{E}$-$\bf{B}$ formulations) and the Newton-Lorentz equations. Therefore, the fundamentals of PIC codes consist of four kernel parts: force depositing to particles, particle pushing, particles depositing to charge and current densities, and solving Maxwell equations; see Figure.~\ref{fig-loop}. Here, we review each part briefly (these algorithms are used in the SLIPs) and refer to the standard literature or textbooks for more details \cite{Arber_2015,Birdsall_2018}.

\subsection{Particle pushing}
When radiation reaction is weak (the radiation power is much smaller than the energy gain power), the motion of charged particles is governed by the Newton-Lorentz equation:
\begin{eqnarray}
	\frac{d{\mathbf{p}}}{d t} &=& \frac{q}{m}(\mathbf{E} + \boldsymbol{\beta}\times\mathbf{B}), \\
	\frac{d\mathbf{x}}{d t} &=& \frac{\mathbf{p}}{\gamma},
\end{eqnarray}
where $\mathbf{p} \equiv \gamma m \mathbf{v}$, $\mathbf{x}$, $q$, $m$, $\gamma$, $\mathbf{v}$, and $\boldsymbol{\beta} \equiv \mathbf{v}/c$ are the momentum, position, charge, mass, Lorentz factor, velocity, and normalized velocity of the particle, respectively. These coupled equations are discretized using a leapfrog algorithm as
\begin{eqnarray}
	\frac{\mathbf{p}^{n+1/2} - \mathbf{p}^{n-1/2}}{\Delta t} &=& \frac{q}{m}\left(\mathbf{E}^n + \frac{\mathbf{p}^n}{\gamma^n} \times \mathbf{B}^n\right), \label{lorentz1} \\
	\frac{\mathbf{x}^{n+1} - \mathbf{x}^n}{\Delta t} &=& \mathbf{v}^{n+1/2}, \label{lorentz2}
\end{eqnarray}
and solved using the standard Boris rotation \cite{Buneman_1967,Boris_1971,Qin_2013}:
\begin{eqnarray}
	\mathbf{p}^{n-1/2} &=& \mathbf{p}^- - \frac{q\Delta t}{2m}\mathbf{E}^n, \label{boris1} \\
	\mathbf{p}^{n+1/2} &=& \mathbf{p}^+ + \frac{q\Delta t}{2m}\mathbf{E}^n, \\
	\mathbf{p}' &=& \mathbf{p}^- + \mathbf{p}^- \times \boldsymbol{\tau}, \\
	\mathbf{p}^+ &=& \mathbf{p}^- + \mathbf{p}' \times \boldsymbol{\varsigma }, \\
	\boldsymbol{\tau} &=& \frac{q\Delta t}{2m\gamma^n}\mathbf{B}^n, \\
	\boldsymbol{\varsigma } &=& \frac{2\boldsymbol{\tau}}{1+\boldsymbol{\tau}^2}, \label{boris6}
\end{eqnarray}
where $\gamma^n = \sqrt{1+(\mathbf{p}^-)^{2}}=\sqrt{1+(\mathbf{p}^+)^{2}}$. The update in momentum and position are asynchronized by half a time step, i.e., a leapfrog algorithm is used here. This leapfrog algorithm ensures the self-consistency of the momentum and position evolution.

\subsection{Field solving} 
In the ultraintense laser-plasma interactions, the plasma particles are assumed to be distributed in the vacuum and immersed in the EM field. Therefore, the field evolution is governed by the Maxwell equations in vacuum with sources. After normalization, the Maxwell equations in differential form are given by 

\begin{eqnarray}
	\nabla \cdot \mathbf{E} &=& \rho \\
	\nabla \cdot \mathbf{B} &=& 0 \\
	\nabla \times \mathbf{E} &=& -\frac{\partial \mathbf{B}}{\partial t} \\
	\nabla \times \mathbf{B} &=& \frac{\partial \mathbf{E}}{\partial t} + \mathbf{J}. \label{ampere}
\end{eqnarray}

\begin{figure}[ht] \centering \includegraphics[width=0.8\linewidth]{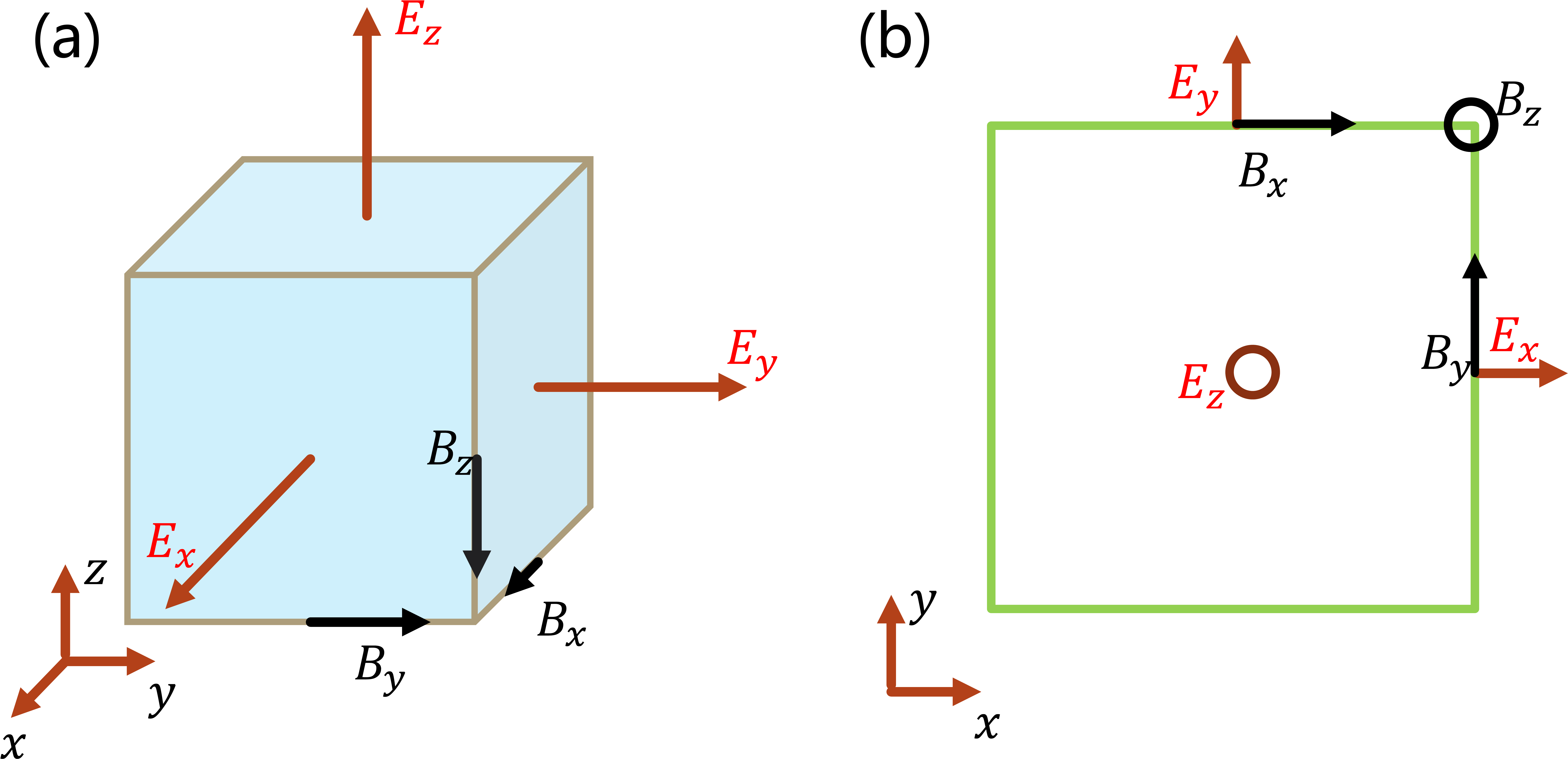} \caption{(a) and (b): Yee-grid and position of each field component in 3D and 2D case, respectively. In (b), the $z$ direction is squeezed.} \label{yeegrid} \end{figure}

The standard finite-difference method in the time domain for the Maxwell equations is to discretize field variables on the spatial grid and advance forward in time. Here, following the well-known Yee-grid \cite{kane_yee_numerical_1966}, we put $\bf{E}, \bf{B}$ as in Figure.~\ref{yeegrid} (a), which automatically satisfies the two curl equations. For lower-dimension simulations, extra dimensions are squeezed, as shown in a 2D example in Figure.~\ref{yeegrid} (b). In the dimension-reduced simulations, field components in the disappeared dimension can be seen as uniform, i.e., the gradient is 0.

By using Esierkepov's method of current deposition \cite{esirkepov_exact_2001}, the current is calculated from the charge density via charge conservation, i.e., $\partial_t \rho + \nabla \cdot {\bf J} = 0$. Once the initial condition obeys Gauss's law, $\nabla \cdot {\bf E} = \rho$, Gauss's law is automatically embedded. This can be verified with a gradient on Eq.~(\ref{ampere}) $0 = \nabla \cdot (\nabla \times {\bf B}) = \partial_t(\nabla \cdot {\bf E}) + \nabla \cdot {\bf J} = \partial_t(\nabla \cdot {\bf E} - \rho)$, i.e., the temporal variation in the violation of Gauss's law is 0. Therefore, in the field solver, only the two curl equations are solved. Here, we take the $E_y$ and $B_z$ components as examples: \\ 
1D case (squeezing the $y,z$ directions): \\ 
\begin{eqnarray}
	\begin{split}
	\frac{E_y^{n+1}-E_y^n}{\Delta t}\bigg|_{i+1/2} =& -\frac{B_{i+1} - B_{i}}{\Delta x}\bigg|_z^{n+1/2}-J_{y,i+1/2}^{n+1/2} \\
	\frac{B_z^{n+1/2}-B_z^{n - 1/2}}{\Delta t}\bigg|_{i} =& -\frac{E_{i+1/2} - E_{i-1/2}}{\Delta x}\bigg|^{n}_{y}
	\end{split}
\end{eqnarray}
2D case (squeezing the $z$ direction): \\ 
\begin{eqnarray}
	\begin{split}
	\frac{E_y^{n+1}-E_y^n}{\Delta t}\bigg|_{i+1/2, j} = & -\frac{B_{i+1, j} - B_{i, j}}{\Delta x}\bigg|_z^{n+1/2} - J^{n+1/2}_{y, i+1/2, j} \\
	\frac{B_z^{n+1/2}-B_z^{n-1/2}}{\Delta t}\bigg|_{i+1/2, j} = & -\frac{E_{i+1/2, j} - E_{i+1/2, j}}{\Delta x}\bigg|_y^{n} + \\ & \frac{E_{i+1/2,j+1/2} - E_{i+1/2,j-1/2}}{\Delta y}\bigg|_x^{n+1/2}
	\end{split}
\end{eqnarray}
3D case: \\ 
\begin{eqnarray}
	\begin{split}
	\frac{E_y^{n+1}-E_y^n}{\Delta t}\bigg|_{i+1/2, j, k+1/2} = & -\frac{B_{i+1, j, k + 1/2} - B_{i, j, k + 1/2}}{\Delta x}\bigg|_z^{n+1/2} + \\ & \frac{B_{i+1/2,j, k+1} - B_{i+1/2,j, k}}{\Delta z}\bigg|_x^{n+1/2}-J_{y,i+1/2,j, k+1/2}^{n+1/2} \\
	\frac{B_z^{n+1/2}-B_z^{n-1/2}}{\Delta t}\bigg|_{i, j, k+1/2} = & -\frac{E_{i+1/2, j, k} - E_{i-1/2, j, k}}{\Delta x}\bigg|_y^{n} + \\ & \frac{E_{i+1/2,j+1, k} - E_{i+1/2,j, k}}{\Delta y}\bigg|_x^{n}
	\end{split},
\end{eqnarray}
where the lower indices with $i, j, k$ denote the spatial discretization and upper indices with $n$ indicate the time discretization. The time indices are assigned using the leapfrog algorithm; see Sec.~\ref{leapfrog}.

\subsection{Current deposition} 
We calculate the charge current density using Esirkepov’s method, which conserves charge by satisfying the Gauss law \cite{Esirkepov_2001} 
\begin{equation}
	\partial_t \rho + \nabla \cdot \mathbf{J} = 0,
\end{equation}
and removes the need for Coulomb correction \cite{Birdsall_2018}. This algorithm computes the charge density at time step $t - \frac{1}{2}\Delta t$ and $t + \frac{1}{2}\Delta t$ on each grid cell from the particle positions and velocities, i.e., 
\begin{eqnarray}
	\rho^{n+1/2}_{i, j, k} &=& \frac{1}{\Delta V}\sum_r W(\mathbf{x}^n_r + \frac{1}{2} \mathbf{v}^{n+1/2} \Delta t) q_r, \\
	\rho^{n-1/2}_{i, j, k} &=& \frac{1}{\Delta V}\sum_r W(\mathbf{x}^n_r - \frac{1}{2} \mathbf{v}^{n} \Delta t) q_r, \\
	\delta^n \rho &=& \rho^{n+1/2} - \rho^{n-1/2}
\end{eqnarray}
where $r$ denotes the particle index, $|{\bf x}_r - {\bf x}_{i, j, k}| \leq (\Delta x, \Delta y, \Delta z)$, and $\Delta V = \Delta x \Delta y \Delta z$ is the cell volume. We then interpolate the charge density to the current grid to obtain the current density; see Ref.~\cite{Esirkepov_2001} for more details.

\subsection{Force deposition} We deposit the updated field variables from the Maxwell solver to the particles for calculating acceleration or further SF-QED processes. The field deposition to the particles follows a similar procedure as the charge density deposition. For each particle at position $\mathbf{x}_r$, we find its nearest grid point $(i, j, k)^g = \mathrm{floor}\left(\frac{\mathbf{x}_r}{ \Delta \mathbf{x}} + \frac{1}{2}\right)$ and its nearest half grid point $(i, j, k)^h = \mathrm{floor}\left(\frac{\mathbf{x}_r}{ \Delta \mathbf{x}}\right)$, where $\Delta \mathbf{x} = (\Delta x, \Delta y, \Delta z)$ is the spatial grid size. We then weight the field to the particle by summing over all nontrivial terms of $W(i, j, k) \cdot F(i, j, k)$, where $W(i, j, k)$ is the particle weighting function (see Sec.~\ref{particlesf} for more details) on the grid (half grid) $(i, j, k)$ and $F(i, j, k)$ is the field component of $\bf{E}$ or $\bf{B}$ on the spatial grid with proper staggering according to Figure.~\ref{yeegrid}.

\subsection{Particle shape function} \label{particlesf} The weighting function $W$ in the current and force deposition is determined by the form factor (shape factor) of the macro-particle, which is a key concept in modern PIC code algorithms. The form factor gives the macro-particle a finite size (composed of thousands of real particles) and reduces the nonphysical collisions \cite{Birdsall_2018}. Various particle shape function models have been proposed, such as the Nearest Grid Point (NGP) and Cloud-in-Cell (CIC) methods. The NGP and CIC methods use the nearest one and two grid fields as the full contribution, respectively. Higher orders of particle shape function can suppress the unphysical noise and produce smoother results. We use the triangle shape function (triangular shape cloud, TSC) in each dimension \cite{Esirkepov_2001} 
\begin{eqnarray}
	W_{\rm spline} = \Bigg\{\begin{array}{ll}
		\frac{3}{4}-{\delta}^2,& \mathrm{for } j \\
		\frac{1}{2}\left(\frac{1}{2} \pm \delta \right),  & \mathrm{for} j\pm1
	\end{array},
\end{eqnarray}
where $\delta = \frac{x - X_j}{\Delta x}$, $x$ is the particle position, $j$ the nearest grid/half grid number and $X_j \equiv j \Delta x$. We obtain higher dimension functions by multiplying 1D shape function in each dimension: $W_\mathrm{2D}(i, j) = W_x(i) W_y(j)$ and $W_\mathrm{3D}(i, j, k) = W_x(i) W_y(j) W_z(k)$.

\subsection{Time ordering} \label{leapfrog}

\begin{figure}[ht] 
	\centering 
	\includegraphics[width=0.5\linewidth]{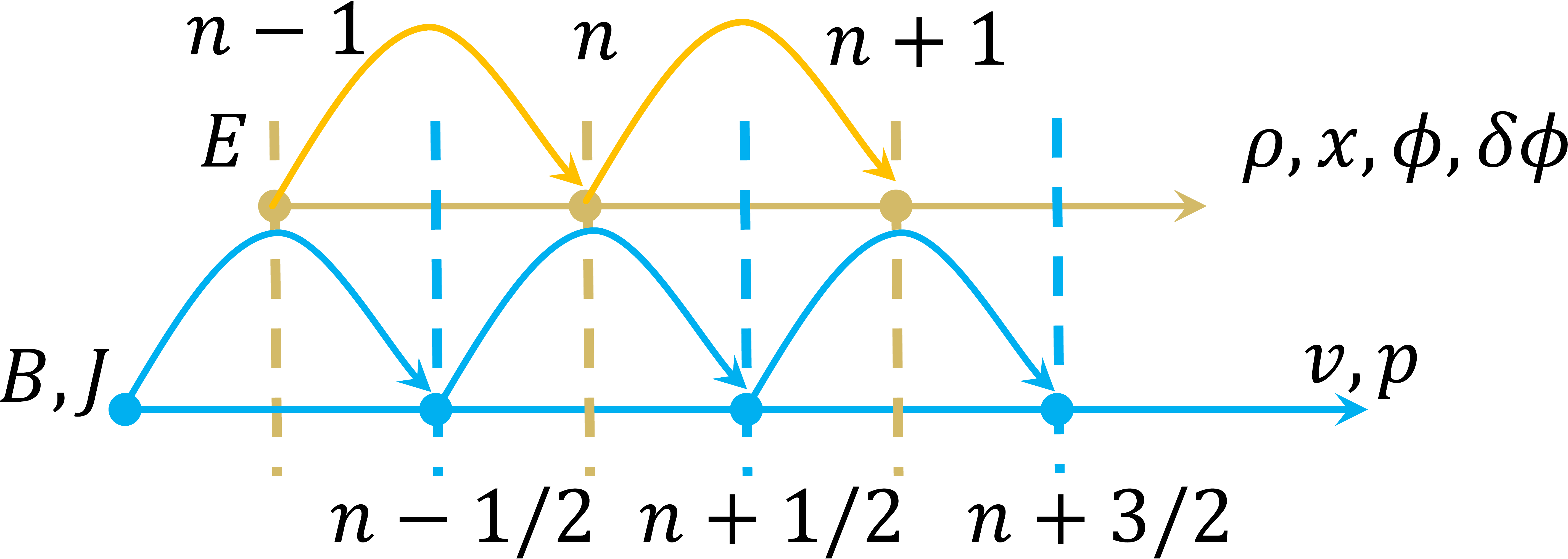} 
	\caption{Leap-frog algorithm of particle pushing and field advancing.} 
	\label{leapfrog-algorithm} 
\end{figure}

In SLIPs, the simplest forward method is used to discretize all differential equations that are reduced to first order with respect to time \cite{Arber_2015}. To minimize the errors introduced by the discretization, some variables are updated at integer time steps and others at half-integer time steps. For example, the EM field variables $\bf{E}$ and $\bf{B}$ are updated alternately at integer and half-integer time steps; and the position $\bf{x}$ and momentum $\bf{p}$ of particles are updated alternately as well; see Figure.~\ref{leapfrog-algorithm}. The leapfrog updating is also applied to the current deposition and field interpolation.

\section{QED Algorithm} 
This section presents some SF-QED processes (with unpolarized/polarized version) that are relevant for laser-plasma interactions. The classical and quantum radiation correction to the Newton-Lorentz equations, namely the Laudau-Lifshitz (LL) equation and the modified Landau-Lifshitz (MLL) equation; and their discretized algorithms are reviewed first. The classical and quantum corrected equations of motion (EOM) for the spin, namely the Thomas-Bargmann-Michel-Telegdi (T-BMT) equation and the Radiative T-BMT equation; and their discretized algorithms are reviewed next. NCS with unpolarized and polarized version and their Monte-Carlo (MC) algorithms are reviewed. NBW with unpolarized and polarized version and their MC implementations are presented as well. Finally, the implementations of high-energy bremsstralung and vacuum birefringence in the conditions of weak pair productions ($\chi_\gamma \lesssim 0.1$) are briefly discussed.

\subsection{Radiative particle pusher} 
Charged particles moving in strong fields can emit either classical fields or quantum photons. This leads to energy/momentum loss and braking of the particles, i.e., radiation reaction. 
{A well-known radiative equation of motion (EOM) for charged particles is the Lorentz-Abraham-Dirac (LAD) equation \cite{LADeqn}. However, this equation suffers from the runaway problem, as the radiation reaction terms involve the derivative of the acceleration. To overcome this issue, several alternative formalisms have been proposed, among which the Landau-Lifshitz (LL) version is widely adopted \cite{Landau_1999_classical}. The LL equation can be obtained from the LAD equation by applying iterative and order-reduction procedures \cite{Ekman2021,Ekman2022}, which are valid when the radiation force is much smaller than the Lorentz force. More importantly, in the limit of $\hbar \rightarrow 0$, the QED results in the planewave background field are consistent with both the LAD equation and LL equation \cite{Ilderton2013Radiation,seipt2023kinetic}. Depending on the value of the quantum nonlinear parameter $\chi_e$ (defined in Sec. \ref{lleqns}), the particle dynamics can be governed by either the LL equation or its quantum-corrected version \cite{Piazza_2012_Extremely, Landau_1999_classical, Neitz_2014_Electron, Derouillat_2018}.
}

\subsubsection{Landau-Lifshitz equation}\label{lleqns}
The Landau-Lifshitz equation can be employed when the radiation is relatively weak (weak radiation reaction, $\chi_e \ll 10^{-2}$) \cite{Landau_1999_classical}:
\begin{widetext}
	\begin{equation}\label{eq1}
		\begin{split}
			\mathbf{F}_{\mathrm{RR, classical}}& =\frac{2e^3}{3mc^3} \bigg \lbrace \gamma \bigg[ \bigg (\frac{\partial}{\partial t}+\frac{\mathbf{p}}{\gamma m}\cdot\nabla \bigg )\mathbf{E}+\frac{\mathbf{p}}{\gamma mc} \times \bigg (\frac{\partial}{\partial t}+\frac{\mathbf{p}}{\gamma m}\cdot\nabla \bigg )\mathbf{B} \bigg] \\
			& + \frac{e}{mc} \bigg[\mathbf{E}\times \mathbf{B}+\frac{1}{\gamma mc}\mathbf{B}\times (\mathbf{B}\times \mathbf{p})+\frac{1}{\gamma mc}\mathbf{E}(\mathbf{p \cdot E}) \bigg] \\ & -\frac{e\gamma}{m^2c^2}\mathbf{p} \bigg [ \bigg(\mathbf{E}+\frac{\mathbf{p}}{\gamma mc}\times \mathbf{B} \bigg)^2-\frac{1}{\gamma^2m^2c^2}(\mathbf{E  \cdot p})^2 \bigg ] \bigg\rbrace,
		\end{split}
	\end{equation}
\end{widetext}
where all quantities are given in Gaussian units, and the dimensionless one is 
\begin{widetext}
	\begin{equation}\label{eq1-1}
		\begin{split}
			\mathbf{F}_{\mathrm{RR,classical}}& =\frac{2}{3}\alpha_f \xi_L \bigg \lbrace \gamma \bigg[ \bigg (\frac{\partial}{\partial t}+\frac{\mathbf{p}}{\gamma}\cdot\nabla \bigg )\mathbf{E}+\frac{\mathbf{p}}{\gamma} \times \bigg (\frac{\partial}{\partial t}+\frac{\mathbf{p}}{\gamma}\cdot\nabla \bigg )\mathbf{B} \bigg] \\
			& + \bigg[\mathbf{E}\times \mathbf{B}+\frac{1}{\gamma}\mathbf{B}\times (\mathbf{B}\times \mathbf{p})+\mathbf{E}(\mathbf{p \cdot E}) \bigg] \\ & -\gamma\mathbf{p} \bigg [ \bigg(\mathbf{E}+\frac{\mathbf{p}}{\gamma}\times \mathbf{B} \bigg)^2-\frac{1}{\gamma^2}(\mathbf{E  \cdot p})^2 \bigg ] \bigg\rbrace,
		\end{split}
	\end{equation}
\end{widetext}
with $\alpha_f = \frac{e^2}{c\hbar}$ is the fine structure constant and $\xi_L = \frac{\hbar\omega}{m_ec^2}$ is the normalized reference photon energy.
In the ultra-intense laser interacting with plasmas, the dominant contribution comes from the last two terms \cite{Tamburini_2010}.
In the ultrarelativisitc limits, only the third term dominates the contribution, and the radiation reaction force can be simplified as
\begin{equation}
	\mathbf{F}_\mathrm{RR, cl} \simeq \frac{2}{3}\alpha_f \frac{\chi^2_e}{\xi_L}\boldsymbol{\beta},
\end{equation}
where $\chi_e = \frac{e\hbar}{m^3 c^4}\sqrt{|F^{\mu \nu}p_\nu|^2} \equiv  \xi_L \gamma_e \sqrt{(\mathbf{E}+\boldsymbol{\beta}\times\mathbf{B})^2 - (\boldsymbol{\beta}\cdot (\boldsymbol{\beta} \cdot \mathbf{E}))^2} \simeq \gamma_e E_\perp \xi_L (1-\cos\theta)$ is a nonlinear quantum parameter signifies the strength of the NCS, with $\theta$ denote the angle between electron momentum and EM field wavevector, and $E_\perp$ denotes the perpendicular component of electric field.
This reduced form gives the importance of the radiation reaction when one estimates the ratio between $F_\mathrm{RR}$ and Lorentz force $F_L$:
\begin{equation}
	\mathcal{R} \equiv |F_\mathrm{RR}|/|F_\mathrm{L}| \sim \frac{2}{3}\alpha_f \gamma_e \chi_e \simeq 2\times10^{-8}a_0\gamma_e^2~(\mathrm{for~wavelength = 1~\mu m}),
\end{equation}
apparently, once $\gamma_e^2 a_0  \gtrsim 10^6$, the radiation reaction force should be considered.

\subsubsection{Modified Landau-Lifshitz equation}
\begin{figure}[h]
	\centering
	\includegraphics[width=0.55\linewidth]{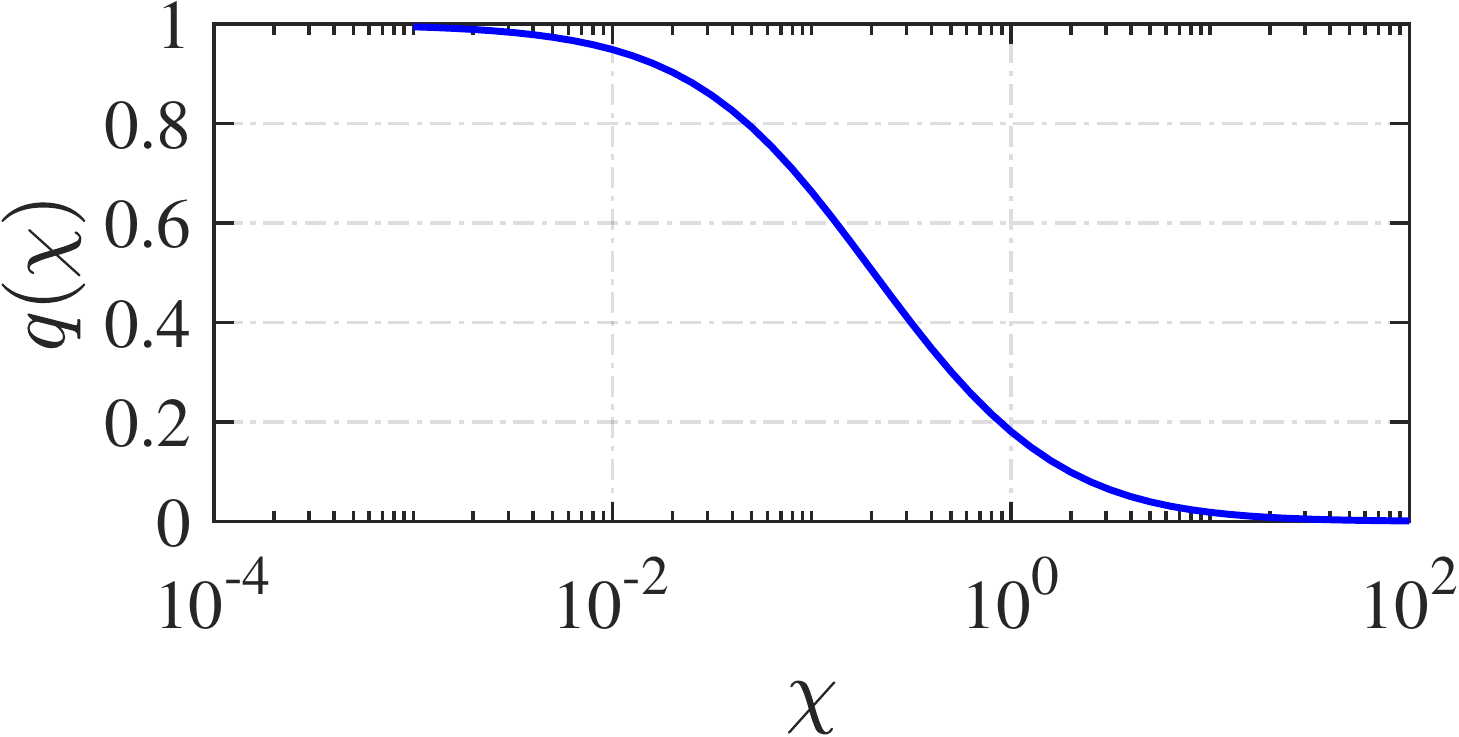}
	\caption{$q(\chi)$ vs $\chi$.} \label{qchi}
\end{figure}

The LL equation is only applicable when the radiation reaction force is much weaker than the Lorentz force, or, the radiation per laser period is much smaller than $m_ec^2$ \cite{Bulanov_2013}.
Once $\chi_e$ is larger than $10^{-2}$ and above, the quantum nature of the radiation dominates the process.
On one hand, the radiation spectra will be suppressed and deviate from the radiation force in LL equation; on the other hand, the radiation will be stochastic and discontinuous.
However, when the stochasticity is not relevance for the detection and one only cares about the average effect (integrated spectra), a correction to the radiation force can be made, i.e., quantum correction \cite{Nikishov_1964_Quantum, Sokolov_2009_Dynamics, Piazza_2010_Quantum, Thomas_2012_Strong}
\begin{equation}
	\mathbf{F}_\mathrm{RR, quantum}=q(\chi)\mathbf{F}_\mathrm{RR, classical},
\end{equation}
where
\begin{eqnarray}
	q(\chi) 	&=& \frac{I_\mathrm{QED}}{I_\mathrm{C}}, \\
	I_\mathrm{QED} &=& mc^2\int c(k\cdot k')\frac{dW_{fi}}{d\eta d r_0}dr_0, \\
	I_\mathrm{C} &=& \frac{2e^4E'^2}{3m^2c^3},
\end{eqnarray}
with $W_{fi}$ being the radiation probability \cite{Sokolov_2010_Emission}, $\eta = k_0 z - \omega_0 t$, $r_0=\frac{2\left(k\cdot k'\right)}{3\chi\left(k\cdot p_i\right)}$, and $E'$ the electric fields in the instantaneous frame of the electron. $p_i$ is the four-momentum of the electron before radiation. $k$ and $k'$ are the four-wavevector of local EM field, and the radiated photon, respectively. See $q(\chi)$ in Figure.~\ref{qchi}. Here, the ratio between the QED radiation power and the classical one, i.e., the re-scaling factor $q(\chi)$, is the same with the factor in Ref.~\cite{Bulanov_2013_extreme}:
\begin{equation}
	q(\chi_e) \approx \frac{1}{\left[1+4.8(1+\chi_e)\ln(1+1.7\chi_e)+2.44\chi_e^2\right]^{2/3}},
\end{equation}
or
\begin{equation}
	q(\chi_e) \approx \frac{1}{(1+8.93\chi_e + 2.41\chi_e^2)^{2/3}}.
\end{equation}

In the ultrarelativistic limit, an alternative formula can be employed as \cite{Niel2018, smilei}
\begin{equation}
	\mathbf{F}_\mathrm{RR,quantum}=q(\chi)P_\mathrm{cl}\chi^2_e \boldsymbol{\beta}/\boldsymbol{\beta}^2c.
\end{equation}
Apparently, once $\chi \gtrsim 10^{-2}$, the quantum corrected version should be used.

\subsubsection{Algorithms of the Radiative Pusher}
Here, we plug the radiative correction (either classical or quantum corrected version) into the standard Boris pusher as follows \cite{Tamburini_2010}:
\begin{equation}
	\frac{\mathbf{p}^{n+1/2}-\mathbf{p}^{n-1/2}}{\Delta t} = \mathbf{F}^n=\mathbf{F}^n_L + \mathbf{F}^n_R.
\end{equation}
First we use the Boris step
\begin{equation}
	\frac{\mathbf{p}_L^{n+1/2} - \mathbf{p}_L^{n-1/2}}{\Delta t} = \mathbf{F}_L^n,
\end{equation}
and then use the radiative correction push
\begin{equation}
	\frac{\mathbf{p}^{n+1/2}_R - \mathbf{p}^{n-1/2}_R}{\Delta t} = \mathbf{F}_R^n,
\end{equation}
where $\mathbf{p}^{n-1/2}_L = \mathbf{p}^{n-1/2}_R = \mathbf{p}^{n-1/2}$, and the final momentum is given by
\begin{equation}
	\mathbf{p}^{n+1/2} = \mathbf{p}^{n+1/2}_L + \mathbf{p}^{n+1/2}_R - \mathbf{p}^{n-1/2} = \mathbf{p}^{n+1/2}_L + \mathbf{F}_R^n \Delta t.
\end{equation}
With this algorithm, the Boris pusher is attained. 

\begin{figure}[ht]
	\centering
	\includegraphics[width=\linewidth]{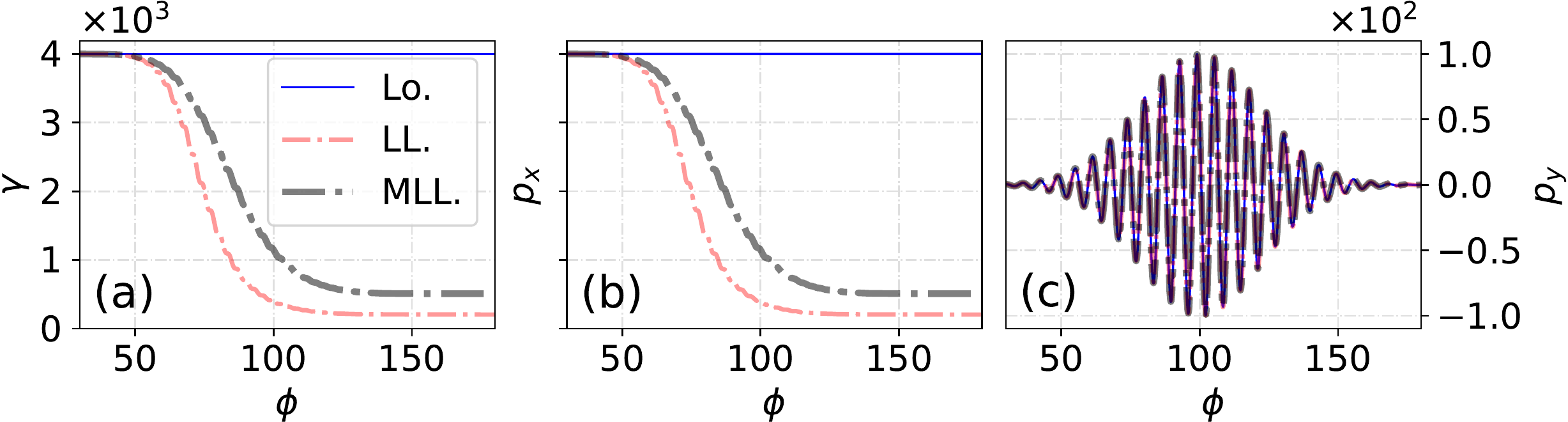}
	\caption{Dynamics of an electron ($\mathbf{p}_0=(4000, 0, 0)$) scattering with an ultraintense linearly polarized laser pulse of $E_y = 100 \exp\left[-\left(\frac{\phi - 100}{10\pi}\right)^2\right]\cos\phi$ with $\phi \equiv t + x$. Here, ``Lo.'', ``LL.'', and ``MLL.'' denote results calculated from Lorentz, LL and modified LL equations, respectively.} \label{pusher}
\end{figure}

See a comparison between different solver calculated dynamics in Figure.~\ref{pusher}.
{
For the Lorentz equation without radiation, the particle momentum and energy are analytically given by \cite{avetissian2015relativistic}
\begin{eqnarray}
	{\bf p}(\tau) &=& {\bf p}_0 - {\bf A}(\tau) + \hat{k} \frac{{\bf A}^2(\tau)-2{\bf p}_0 \cdot {\bf A}(\tau)}{2(\gamma_0 - {\bf p}_0 \cdot \hat{k})} \\
	\gamma(\tau) &=& \gamma_0 + \frac{{\bf A}^2(\tau)-2{\bf p}_0 \cdot {\bf A}(\tau)}{2(\gamma_0 - {\bf p}_0 \cdot \hat{k})} 
\end{eqnarray}
where ${\bf A}(\tau) = - \int{\tau_0}^\tau {\bf E(\tau)} d \tau $ is the external field vector potential, $\tau$ is the proper time, $\hat{k}$ is the normalized wavevector of the field, $\gamma,{\bf p}$, and $\gamma_0,{\bf p}_0$ the instantaneous and initial (with notation of $0$) Lorentz factor and momentum of the particle, respectively. 
For a planewave with a temporal profile, the momentum and energy gain vanish as ${\bf A}(\infty) = {\bf A}(-\infty) = 0$.
The planewave solution with radiation reaction can be found in Ref.~\cite{Piazza2008}.
However, no explicit solution exists when the quantum correction term is included, as shown in Fig.~\ref{pusher}.
}

\subsection{Spin dynamics}
The consideration of electron/positron spin becomes crucial in addition to the kinetics when plasma electrons are polarized or when there is an ultrastrong EM field interacting with electron/positron and $\gamma$ photons. The significance of this aspect has been highlighted in recent literature, particularly in the context of relativistic charged particles in EM waves and laser-matter interactions \cite{Hu_1999, Walser_2002}. This issue can be addressed either by employing the computational Dirac solver \cite{Mocken_2008} or by utilizing the Foldy-Wouthuysen transformation and the quantum operator formalism, such as the reduction of the Heisenberg equation to a classical precession equation \cite{Heiko_2014, Wen_2017}. However, these approaches are not directly applicable to many-particle systems.
Here and throughout this paper, the spin is defined as a unit vector $\mathbf{S}$. In the absence of radiation, the electron/positron spin precesses around the magnetic field in the rest frame and can be described by the classical Thomas-Bargmann-Michel-Telegdi (T-BMT) equation. This equation is equivalent to the quantum-mechanical Heisenberg equation of motion for the spin operator or the polarization vector of the system \cite{Heiko_2014, Wen_2017, Jackson_2021}.
When radiation becomes significant, the electron/positron spin also undergoes flipping to quantized axes, typically aligned with the magnetic field in the rest frame. By neglecting stochasticity, this effect can be appropriately accounted for by incorporating the radiative correction to the T-BMT equation, which is analogous to the quantum correction to the Landau-Lifshitz equation.

\subsubsection{T-BMT equation}
The non-radiative spin dynamics of an electron is given by
\begin{equation}
	\begin{split}
	\left(\frac{{\rm d}{\bf S}}{{\rm d}t}\right)_{T}= ~&\mathbf{S} \times \mathbf{\Omega} \equiv {\bf S}\times \left[-\left(\frac{g}{2}-1\right)\frac{\gamma_e}{\gamma_e+1}\left({\boldsymbol \beta}\cdot{\bf B}\right)\cdot{\bm \beta} \right.\\ &\left. +\left(\frac{g}{2}-1+\frac{1}{\gamma_e}\right){\bf B}-\left(\frac{g}{2}-\frac{\gamma_e}{\gamma_e+1}\right)\boldsymbol{\beta}\times{\bf E}\right],
	\end{split}\label{bmteqn}
\end{equation}
where ${\bf E}$ and ${\bf B}$ are the normalized electric and magnetic fields, respectively, and $g$ is the electron Land\'e factor, respectively. Since this equation is a pure rotation around the precession frequency of $\mathbf{\Omega}$, the Boris rotation is much more preferable than other solver for ordinary differential equations (for instance, the Runge-Kutta, etc.). Here, $\mathbf{\Omega}$ plays the role of $\mathbf{B}/\gamma$ in the Eqs.~(\ref{lorentz1}) and (\ref{boris1}-\ref{boris6}). For other particle species, both the charge, mass and Land\'e factor for that species should be employed.

\subsubsection{Radiative T-BMT equation}
When the radiation damping is no longer negligible, the radiation can also affect the spin dynamics. In the weakly radiation regime, this radiation induced modification of the spin dynamics can be handled with a similar way as in the Landau-Lifshitz equation. Here, the modified version of T-BMT equation or the radiative T-BMT equation is thus given by
\begin{equation}\label{spin1}
	\frac{{\rm d}{\bf S}}{{\rm d}t}=\left(\frac{{\rm d}{\bf S}}{{\rm d}t}\right)_{T}+\left(\frac{{\rm d}{\bf S}}{{\rm d}t}\right)_{R},
\end{equation}
with the first (labeled with ``T'') and second (labeled with ``R'') terms denote the non-radiative precession in Eq.~(\ref{bmteqn}), and radiative correction, respectively. The radiative term is given by
\begin{equation}
	\left(\frac{{\rm d}{\bf S}}{{\rm d}t}\right)_{R}=-P\left[\psi_1(\chi){\bf S}+\psi_2(\chi)({\bf S}\cdot{\boldsymbol \beta}){\boldsymbol \beta}+\psi_3(\chi){\hat{\bf n}}_B\right],
\end{equation}
where, $P= \frac{\alpha_f}{\sqrt{3}\pi \gamma_e \xi_L}$, $\psi_1(\chi_e)$ = $\int_{0}^{\infty} u''{\rm d}u {\rm K}_{\frac{2}{3}}(u')$, $\psi_2(\chi_e)$ = $\int_{0}^{\infty} u''{\rm d}u \int_{u'}^{\infty}{\rm d}x{\rm K}_{\frac{1}{3}}(x)$-$\psi_1(\chi_e)$, $\psi_3(\chi)$ = $\int_{0}^{\infty} u''{\rm d}u {\rm K}_{\frac{1}{3}}(u')$,  $u'=2u/3\chi_e$, $u''=u^2/(1+u)^3$, and ${\rm K}_n$ is the $n$th-order modified Bessel function of the second kind,  $\hat{{\bf n}}_B={\boldsymbol \beta}\times\hat{{\bf a}}$, ${\boldsymbol \beta}$ and $\hat{{\bf a}}$ denote the normalized velocity and acceleration vector \cite{Baier_1972, Guo_2020}.

\subsubsection{Algorithms of simulating the spin precession}
The simulation algorithms of spin precession are quite similar to the cases of EOM, i.e., Lorentz equation and radiative EOM, i.e., LL/MLL equations.
Therefore, the T-BMT equation is simulated via the Boris rotation without the pre- and post- acceleration term, but with only the rotation term $\mathbf{\Omega}$. In SLIPs, a standard Boris algorithm is used
\begin{eqnarray}
	\mathbf{S}' &=& \mathbf{S}^{n-1/2} + \mathbf{S}^{n-1/2} \times \mathbf{t}, \label{bmt-boris1} \\
	\mathbf{S}^{n+1/2} &=& \mathbf{S}^{n-1/2} + \mathbf{S}' \times \mathbf{o}, \label{bmt-boris2} \\
	\mathbf{t} &=& \frac{q\Delta t}{2}\mathbf{\Omega}^n, \label{bmt-boris3} \\
	\mathbf{o} &=& \frac{2\mathbf{t}}{1+t^2}. \label{bmt-boris4}
\end{eqnarray}

For the radiative T-BMT equation, there will be an extra term $\left(\frac{d\mathbf{S}}{dt}\right)_R$ which is equivalent to the electric field term in the Lorentz equation. Therefore, the straightforward algorithm is given by
\begin{equation}
	\mathbf{S}^{n-1/2}_T = \mathbf{S}^{n-1/2} + \frac{\Delta t}{2} \left(\frac{d\mathbf{S}}{dt}\right)_R,
\end{equation}
\begin{center}
Boris~T-BMT~Eqs.~(\ref{bmt-boris1}-\ref{bmt-boris4}),
\end{center}
\begin{equation}
	\mathbf{S}^{n+1/2} = \mathbf{S}^{n+1/2}_T + \frac{\Delta t}{2} \left(\frac{d\mathbf{S}}{dt}\right)_R.
\end{equation}

\begin{figure}[ht]
	\centering
	\includegraphics[width=0.8\linewidth]{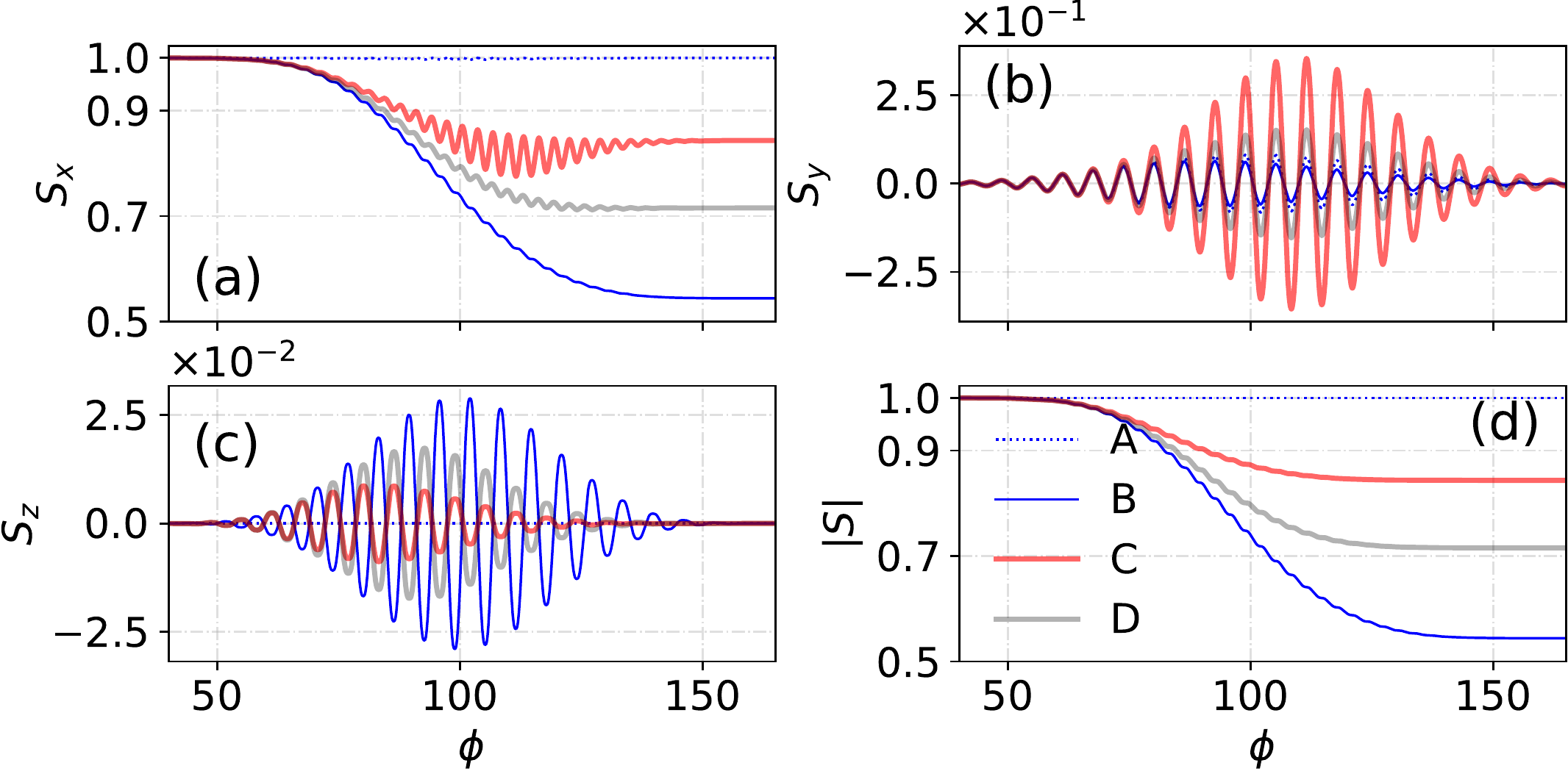}
	\caption{Spin dynamics of an electron [$\mathbf{p}_0=(4000, 0, 0)$, $\mathbf{s}_0 = (1, 0, 0)$] scattering with an ultraintense linearly polarized laser pulse of $E_y = 100 \exp\left[-\left(\frac{\phi - 100}{10\pi}\right)^2\right]\cos\phi$ with $\phi \equiv t + x$. Here, ``A'', ``B'', ``C'', and ``D'' denote calculated via equations of Lorentz + BMT, Lorentz + M-BMT, LL + M-BMT, and MLL + M-BMT.}
	\label{fig-spin}
\end{figure}

{Figure.~\ref{fig-spin} shows the comparison between the T-BMT and radiative T-BMT equations for different cases: Lorentz equation + BMT equation (``A"), Lorentz equation + M-BMT equation (``B"), LL equation + M-BMT equation (``C"), and MLL equation + M-BMT equation (``D"). The evolution of each spin component depends on different terms. In our setup, the magnetic field is along the $z$ direction, so the spin precession occurs in the $x$-$y$ plane, affecting $S_x$ and $S_y$. The radiation reaction mainly affects $S_z$. In the case without radiation reaction, case ``A", $S_x$ and $S_y$ oscillate due to the precession and are conserved in Fig.~\ref{fig-spin}(d). In the case with only the spin radiation reaction, case ``B", $S_x$ is strongly damped by the term $\left(\frac{d{\bf S}}{d t}\right)_R$. $S_y$ and $S_z$ oscillate due to the combined effects of precession and radiation reaction, as shown in Figs.~\ref{fig-spin}(a) and (b). When both spin and momentum radiation reactions are included, case ``C" (LL equation), the particle's momentum and energy decrease, i.e., $\gamma_e$ decreases, which lowers the spin radiation reaction term $\left(\frac{d{\bf S}}{d t}\right)_R(\chi_e)$ and the damping of $S_x$ and $S_z$ (see Fig.~\ref{fig-spin}(c) for the comparison of ``B", ``D" and ``C" in terms of $S_z$ amplitude). Simultaneously, the precession term $\left(\frac{d {\bf S}}{dt}\right)_T \propto B / \gamma_e$ grows with decreasing $\gamma_e$, which amplifies the oscillation of $S_y$, as shown by the contrast of ``B" (Lorentz), ``D" (MLL) and ``C" (LL) in Fig.~\ref{fig-spin}(b).
}

\subsection{Nonlinear Compton scattering}
When the radiation is strong ($\chi_e \gtrsim 0.1$), the stochastic nature of the radiation can no longer be neglected in the laser-beam/plasma interactions. And the photon dynamics should be taken into account. In this regime, the full stochastic quantum process is required to describe the strong radiation, i.e., the nonlinear Compton scattering (NCS) \cite{Bell2008,Kirk_2009_Pair,Ridgers_2014}.
Therefore, the radiation reaction and photon emission process will be calculated via the MC simulation based on the NCS probabilities. Besides, the spin of electron/positron and polarization of the NCS photons will be also included in the MC simulations. 

\subsubsection{Spin-resolved/summed nonlinear Compton scattering}
When the laser intensity $a_0$ and the electron energy $\gamma_e$ permits the local-constant-cross field approximate (LCFA), i.e., $a_0 \gg 1$, $\chi_e \gtrsim 1$, the polarization- and spin-resolved emission rate for the NCS is given by \cite{Li_2020_Polarized,Xue2020,Liu_2022}
\begin{equation}\label{NCSW}
	\frac{{\rm d^2}W_{fi}}{{\rm d}u{\rm d}t}=\frac{W_R}{2}\left(F_0+\xi_1 F_1+\xi_2 F_2 + \xi_3 F_3\right),
\end{equation}
where, the photon polarization is represented by the Stokes parameters  ($\xi_1$, $\xi_2$, $\xi_3$), defined with respect to the axes $\hat{{\bf P}}_1=\hat{\bf a}-\hat{\bf n}(\hat{\bf n}\cdot\hat{\bf a})$ and $\hat{{\bf P}}_2=\hat{\bf n}\times\hat{\bf P}_1$ \cite{Green2015}, with the photon emission direction $\hat{\bf n}={\bf p}_e/|{\bf p}_e|$ along the momentum ${\bf p}_e$ of the ultrarelativistic electron. The variables introduced in Eq.~(\ref{NCSW}) read:
\begin{eqnarray}
	\label{F0}
	F_0&=&-(2+u)^2 \left[{\rm IntK}_{\frac{1}{3}}(u')
	-2{\rm K}_{\frac{2}{3}}(u') \right](1+{S}_{if})+u^2(1-{S}_{if})\nonumber\\
	&&\left[{\rm IntK}_{\frac{1}{3}}(u')
	+2{\rm K}_{\frac{2}{3}}(u') \right]+2u^2{S}_{if}{\rm IntK}_{\frac{1}{3}}(u')-(4u+2u^2)\nonumber\\
	&&({\bf S}_f+{\bf S}_i) \cdot \left[\hat{\bf n}\times\hat{{\bf a}}\right]{\rm K}_{\frac{1}{3}}(u')-2u^2({\bf S}_f-{\bf S}_i) \cdot \left[\hat{\bf n}\times\hat{{\bf a}}\right]{\rm K}_{\frac{1}{3}}(u')\nonumber\\
	&&-4u^2\left[{\rm IntK}_{\frac{1}{3}}(u')
	-{\rm K}_{\frac{2}{3}}(u') \right]({\bf S}_i\cdot\hat{\bf n})({\bf S}_f\cdot\hat{\bf n}),
\end{eqnarray}
\begin{eqnarray}\label{F1}
	F_1&=&-2u^2{\rm IntK}_{\frac{1}{3}}(u')
	\left\{({\bf S}_{i} \cdot \hat{\bf a}){\bf S}_{f} \cdot \left[\hat{\bf n}\times\hat{\bf a}\right]+({\bf S}_{f} \cdot \hat{\bf a}){\bf S}_{i} \cdot \left[\hat{\bf n}\times\hat{\bf a}\right]\right\}+\nonumber\\
	&&4u\left[({\bf S}_i\cdot\hat{\bf a})(1+u)+({\bf S}_f\cdot\hat{\bf a})\right]{\rm K}_{\frac{1}{3}}(u')+\nonumber\\
	&&2u(2+u)\hat{\bf n} \cdot [{\bf S}_f\times{\bf S}_i]{\rm K}_{\frac{2}{3}}(u'),
\end{eqnarray}
\begin{eqnarray}\label{F2}
	F_2&=&-\left\{2u^2 \left\{({\bf S}_{i}\cdot\hat{\bf n}){\bf S}_{f} \cdot \left[\hat{\bf n}\times\hat{\bf a}\right]+({\bf S}_{f}\cdot \hat{\bf n}){\bf S}_{i} \cdot \left[\hat{\bf n}\times\hat{\bf a}\right]\right\}+2u(2+u)\right.\nonumber\\
	&&\left.\hat{\bf a} \cdot [{\bf S}_f\times{\bf S}_i]\right\}{\rm K}_{\frac{1}{3}}(u')-4u\left[({\bf S}_i\cdot\hat{\bf n})+({\bf S}_f\cdot\hat{\bf n})(1+u)\right]\nonumber\\&&{\rm IntK}_{\frac{1}{3}}(u')+4u(2+u)\left[({\bf S}_i\cdot\hat{\bf n})+({\bf S}_f\cdot\hat{\bf n})\right]{\rm K}_{\frac{2}{3}}(u'),
\end{eqnarray}
\begin{eqnarray}\label{F3}
	F_3&=&4\left[1+u+(1+u+\frac{u^2}{2}){S}_{if}-\frac{u^2}{2}({\bf S}_i\cdot\hat{\bf n})({\bf S}_f\cdot\hat{\bf n})\right]{\rm K}_{\frac{2}{3}}(u')\nonumber\\
	&&+2u^2\left\{{\bf S}_i \cdot \left[\hat{\bf n}\times\hat{\bf a}\right]{\bf S}_f \cdot \left[\hat{\bf n}\times\hat{\bf a}\right]-({\bf S}_i\cdot\hat{\bf a})({\bf S}_f\cdot\hat{\bf a})\right\}{\rm IntK}_{\frac{1}{3}}(u')\nonumber\\
	&&-4u\left[(1+u){\bf S}_i\left[\hat{\bf n}\times\hat{\bf a}\right]+{\bf S}_f\left[\hat{\bf n}\times\hat{\bf a}\right]\right]{\rm K}_{\frac{1}{3}}(u'),
\end{eqnarray}
where $W_R=\alpha_f/\left[8\sqrt{3}\pi\xi_L{\left(1+u\right)^3}\right]$, $u'=2u/3\chi$, $u=\omega_\gamma/\left(\varepsilon_i-\omega_\gamma\right)$, ${\rm IntK}_{\frac{1}{3}}(u')\equiv \int_{u'}^{\infty} {\rm d}z {\rm K}_{\frac{1}{3}}(z)$,   $\omega_\gamma$ the emitted photon energy, $\varepsilon_i$ the electron energy before radiation, $\hat{{\bf a}}={\bf a}/|{\bf a}|$ the direction of the electron acceleration ${\bf a}$,
${\bf S}_{i}$ and ${\bf S}_f$ denote the electron spin  vectors before and after radiation, respectively, $|{\bf S}_{i,f}|=1$, and ${S}_{if}\equiv {\bf S}_i\cdot{\bf S}_f$.

Summing over the photon polarization, the electron spin-resolved emission probability can be written as \cite{Li_2020_Polarized, Wan_2020, Xue2020}:
\begin{eqnarray}\label{Wspin}
	&&\frac{{\rm d}^2W_{fi}}{{\rm d}u{\rm d}t}=W_R\left\{-(2+u)^2 \left[{\rm IntK}_{\frac{1}{3}}(u')
	-2{\rm K}_{\frac{2}{3}}(u') \right](1+{S}_{if})+\right.\nonumber\\
	&&\left.	u^2\left[{\rm IntK}_{\frac{1}{3}}(u')
	+2{\rm K}_{\frac{2}{3}}(u') \right](1-{\bf S}_{if})+2u^2{S}_{if}{\rm IntK}_{\frac{1}{3}}(u')-\right.\nonumber\\
	&&\left.(4u+2u^2)({\bf S}_f+{\bf S}_i)\left[{\bf n}\times\hat{{\bf a}}\right]{\rm K}_{\frac{1}{3}}(u')-2u^2({\bf S}_f-{\bf S}_i)\left[{\bf n}\times\hat{{\bf a}}\right]\right.\nonumber\\
	&&\left.	{\rm K}_{\frac{1}{3}}(u')-4u^2\left[{\rm IntK}_{\frac{1}{3}}(u')
	-{\rm K}_{\frac{2}{3}}(u') \right]({\bf S}_i\cdot{\bf n})({\bf S}_f\cdot{\bf n}) \right\}.
\end{eqnarray}

Summing over the final states ${\bf S}_f$,  the initial spin-resolved radiation probability is obtained:
\begin{eqnarray}\label{Wspin2}
	\frac{{\rm d}^2{\overline{W}}_{fi}}{{\rm d}u{\rm d} t}&=&8W_R\left\{-(1+u){\rm IntK}_{\frac{1}{3}}(u')
	+(2+2u+u^2){\rm K}_{\frac{2}{3}}(u')\right.\nonumber\\
	&&\left.-u{\bf S}_i\cdot\left[{\bf n}\times\hat{{\bf a}}\right]{\rm K}_{\frac{1}{3}}(u')\right\}.
\end{eqnarray}
And by averaging the electron initial spin, one obtains the widely used radiation probability for the unpolarized initial particles \cite{Ritus_1979_,Nikishov_1964_Quantum,Ritus_1985_Quantum}.

During the photon emission simulation, electron/positron spin transitions to either a parallel or antiparallel orientation with respect to the spin quantized axis (SQA), depending on the occurrence of emission. Upon photon emission, the SQA  is choosen to obtains the maximum transition probabily, which is along the energy-resolved average polarization 
\begin{eqnarray}\label{S_R}
\mathbf{S}_{\rm f}^{\rm R} = \frac{\mathbf{g}}{(w+\mathbf{f}\cdot\mathbf{S}_i)}.
\end{eqnarray}
These are obtained by summing over the photon polarization and keeps the dependence on initial and final spin of electrons:
\begin{eqnarray}\label{Wspin}
	\frac{{\rm d}^2W_{\rm rad}}{{\rm d}u{\rm d}t}&=&{W_{\rm r}}(w+\mathbf{f}\cdot\mathbf{S}_i+\mathbf{g}\cdot\mathbf{S}_f),
\end{eqnarray}
where
\begin{eqnarray}
	w &=& -(1+u){\rm K}_{\frac{1}{3}}(\rho')+(2+2u+u^2){\rm K}_{\frac{2}{3}}(\rho'), \nonumber\\
	\mathbf{f}&=&u{\rm IntK}_{\frac{1}{3}}(\rho')\hat{{\mathbf v}}\times\hat{{\mathbf a}}, \nonumber\\
	\mathbf{g}&=& -(1+u)\left[{\rm K}_{\frac{1}{3}}(\rho')-2{\rm K}_{\frac{2}{3}}(\rho')\right]\mathbf{S}_i-(1+u)u \nonumber\\
	&&\times{\rm IntK}_{\frac{1}{3}}(\rho')\hat{{\mathbf v}}\times\hat{{\mathbf a}}-u^2\left[{\rm K}_{\frac{1}{3}}(\rho')-{\rm K}_{\frac{2}{3}}(\rho')\right](\mathbf{S}_i\cdot\hat{{\mathbf v}})\hat{{\mathbf v}}. \nonumber
\end{eqnarray}

\begin{figure}[ht]
\centering
\includegraphics[width=\linewidth]{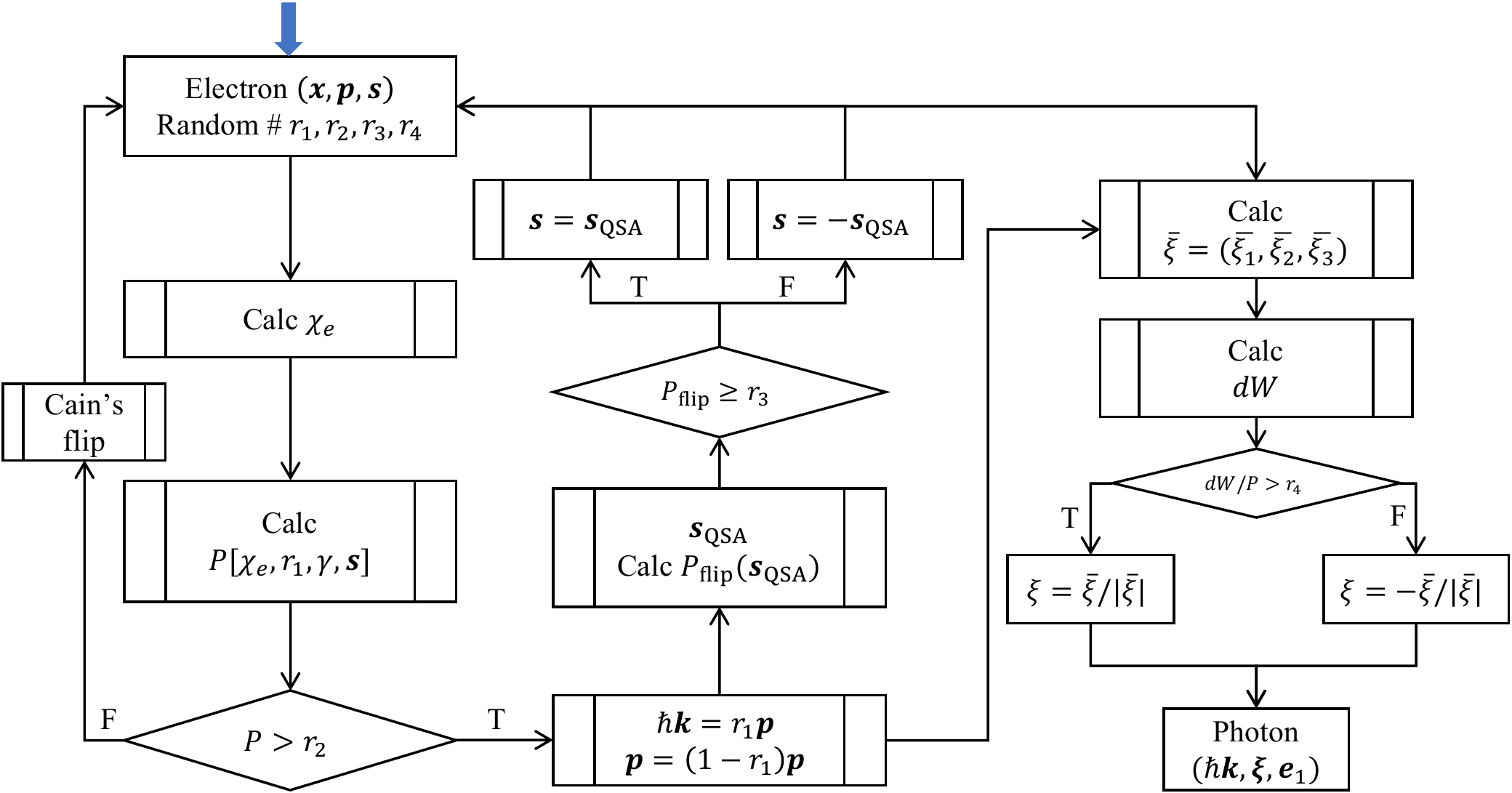}
\caption{Flowchart of the spin- and polarization-resolved NCS}
\label{ncs-loop}
\end{figure}

Conversely, without emission, the SQA aligns with another SQA \cite{Li_2020_Polarized, yokoya2011cain}. In both cases, the final spin is determined by assessing the probability density for alignment, either parallel or antiparallel, with the SQA.
We account for the stochastic spin flip during photon emission using four random numbers $r_{1,2,3,4} \in [0, 1)$. The procedure is as follows: First, at each simulation time step $\Delta t$, a photon with energy $\omega_\gamma = r_1 \gamma_e$ is emitted if the spin-dependent radiation probability in Eq.(\ref{Wspin2}), $P\equiv \mathrm{d}^2\overline{W}_{fi}(\chi_e, r_1, \gamma_e, {\bf S}_i)/\mathrm{d}u \mathrm{d}t \cdot \Delta t$, meets or exceeds $r_2$, following the ``von Neumann's rejection method". The final momentum for the electron and photon is given by ${\bf p}_f = (1 - r_1) {\bf p}_i$ and $\hbar {\bf k} = r_1 {\bf p}_i$, respectively.
Next, the electron spin flips either parallel (spin-up) or antiparallel (spin-down) to the SQA with probabilities of $P_\mathrm{flip} \equiv W_{fi}^{\uparrow}/P$ and $W_{fi}^{\downarrow}/P$, respectively, where $W_{fi}^{\uparrow, \downarrow} \equiv \mathrm{d^2}W_{fi}^{\uparrow, \downarrow}/\mathrm{d}u\mathrm{d}t \cdot \Delta t$ from Eq.~(\ref{Wspin}). In other words, the final spin ${\bf S}_f$ will flip parallel to the SQA if $r_3 < P_\mathrm{flip}$, and vice versa; see the flow chart of the NCS in Figure.~\ref{ncs-loop}.
In the alternative scenario, i.e., no photon is emitted, the average final spin is given by $\overline{\bf S}_f = \frac{{\bf S}_i (1 - W \Delta t) - {\bf f} \Delta t}{1 - (W + {\bf f} \cdot {\bf S}_i)\Delta t}$, where, $W \equiv 16W_R(-(1+u)\mathrm{IntK}_{1/3}(u') + (2+2u+u^2)\mathrm{K}_{2/3}(u'))$, and ${\bf f} \equiv -16W_R({\bf n}\times{\bf \hat{a}}\mathrm{K}_{1/3}(u'))$ \cite{yokoya2011cain,Li_2020_Polarized}. Then the SQA is given by $\overline{\bf S}_f/|\overline{\bf S}_f|$, and the probability for the aligned case is given by $|\overline{\bf S}_f|$, and $1 - |\overline{\bf S}_f|$ for the anti-parallel case.

Finally, the polarization of the emitted photon is determined by considering that the average polarization is in a mixed state. The basis for the emitted photon is chosen as two orthogonal pure states with the Stokes parameters $\hat{\boldsymbol{\xi}}^{\pm}\equiv\pm$($\overline{\xi}_1$, $\overline{\xi}_2$, $\overline{\xi}_3$)/$\overline{\xi}_0$, where $\overline{\xi}_0\equiv \sqrt{(\overline{\xi}_1)^2+(\overline{\xi}_2)^2+(\overline{\xi}_3)^2}$.
The probabilities for the photon emission in these states $W_{fi}^\pm$ are given by Eq.~(\ref{NCSW}). A stochastic procedure is defined using the 4th random number $r_4$: if $W_{fi}^+/\overline{W}_{fi}\geq r_4$, the polarization state $\hat{\boldsymbol{\xi}}^{+}$ will be chosen; otherwise, the polarization state will be assigned as $\hat{\boldsymbol{\xi}}^{-}$. Here, $\overline{W}_{fi}\equiv W_R F_0$ and $W^\pm_{fi} \equiv W_R(F_0 + \sum_{j = 1,3} \xi^\pm_j F_j)$.

Between photon emissions, the electron dynamics in the external laser field are described by the Lorentz equations, $d{\bf p}/dt=-e(\bf{E}+\boldsymbol{\beta}\times\bf{B})$, and are simulated using the Boris rotation method, as shown in Eqs.~(\ref{boris1}-\ref{boris6}). Due to the small emission angle for an ultrarelativistic electron, the photon is assumed to be emitted along the parental electron velocity, i.e.,  ${\bf p}_f \approx (1-\omega{\gamma}/|{\bf p}_i|) {\bf p}i$. 
Besides, in this simulation, interference effects between emissions in adjacent coherent lengths ($l_f \simeq \lambda_L / a_0$) are negligible when the employed laser intensity is ultrastrong, i.e., $a_0\gg 1$. Therefore, the photon emissions happening in each coherent length are independent of each other.

\begin{figure}[ht]
	\centering
	\includegraphics[width=\linewidth]{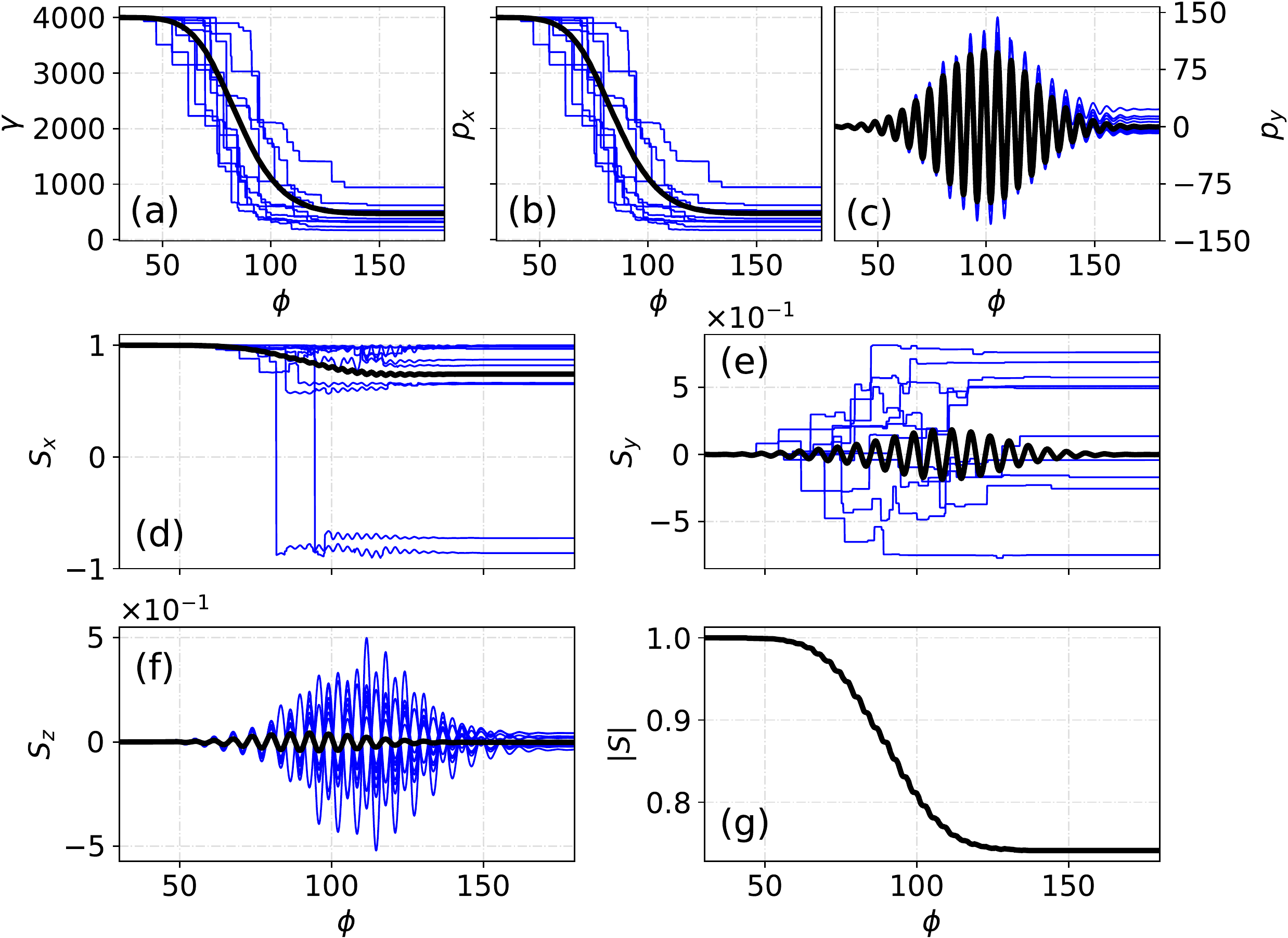}
\caption{Dynamics of 1000 electrons via the stochastic NCS with the simulation parameters as those in Figure.~\ref{fig-spin}. Blue lines are 10 sampled electrons, and black ones are the average value over 1000 sample particles.}
	\label{fig-mc-dynamics}
\end{figure}

\begin{figure}
	\centering
	\includegraphics[width=\linewidth]{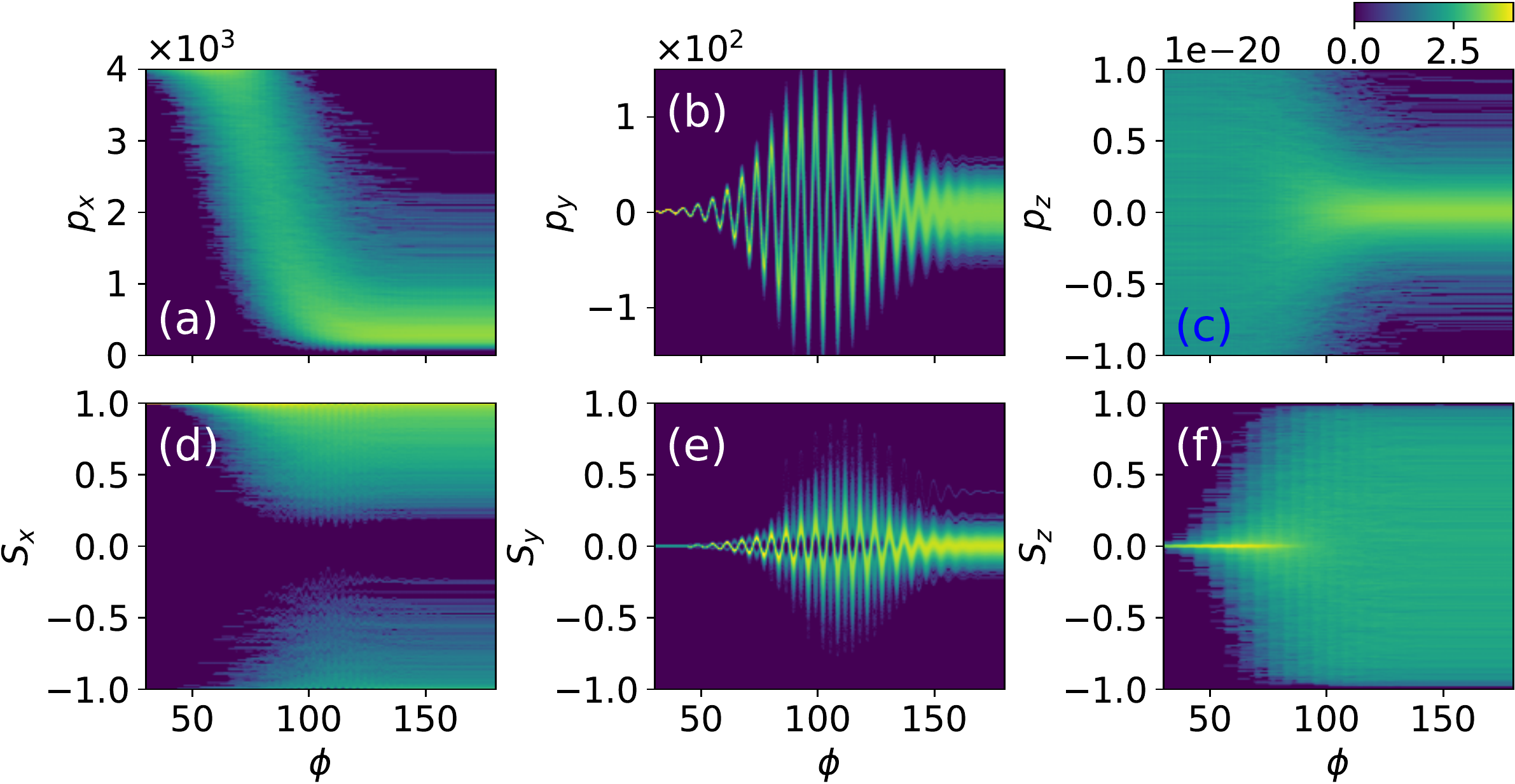}
	\caption{Dynamics of an electron beam (particle number $N_e = 10^4$) with colors denote the number density in arbitrary units and logarithm scale (a. u.), other parameters are the same as those in Figure.~\ref{fig-spin}.} \label{fig-beam-dynamics}
\end{figure}

\begin{figure}[ht]
	\centering
	\includegraphics[width=0.9\linewidth]{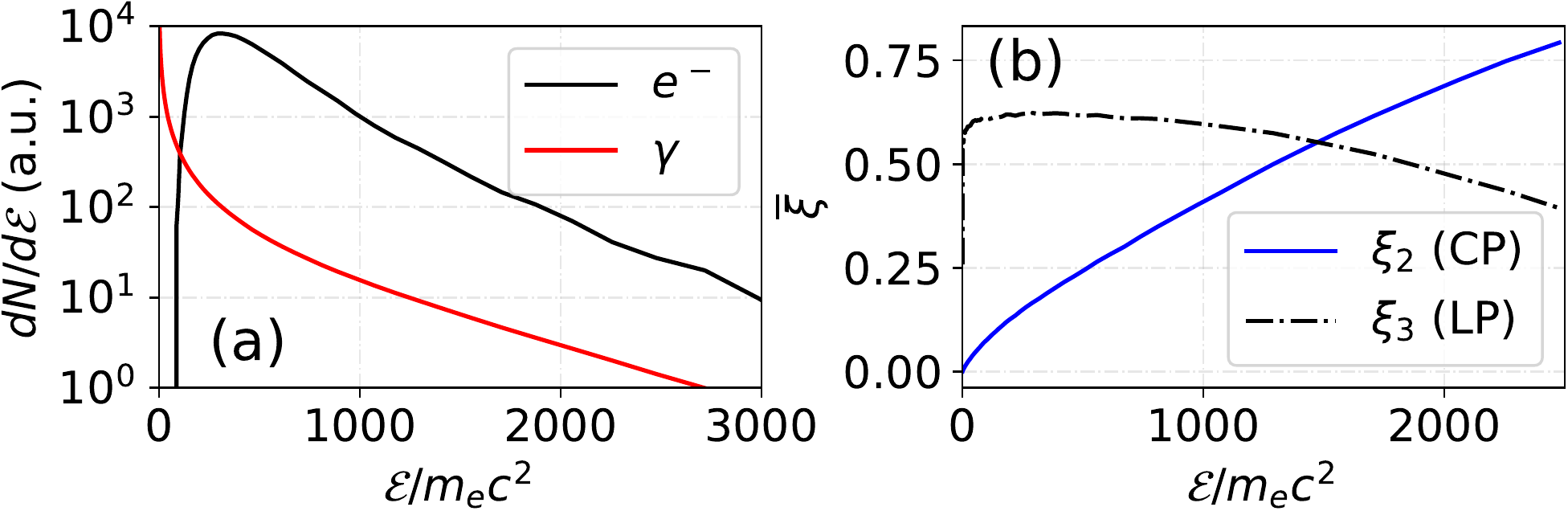}
	\caption{(a) Energy spectra of the scattered electrons (black line)  and generated photons (red line), respectively. (b) Energy-dependent Stokes parameter $\bar{\xi}_2, \bar{\xi_3}$, i.e., circular and linear polarization with respect to $y$-$z$ axis. Simulation parameters are the same as those in Figure.~\ref{fig-spin}.} \label{fig-ncs-spectra}
\end{figure}

Examples of the electron dynamics and spin can be seen in Figure.\ref{fig-mc-dynamics}, apparently, the average value matches the MLL equations for dynamics and MLL + M-BMT equation for spins.  Besides, the beam evolution is shown in Figure.~\ref{fig-beam-dynamics}. The energy spectra of electrons and photons, as well as the photon polarization, can be observed in Figure.~\ref{fig-ncs-spectra}.

\subsubsection{Definition and Transformation of Stokes Parameters}
\label{StokesTra.}

In the context of NCS and the subsequent nonlinear Breit-Wheeler (NBW) pair production, the polarization state of a photon can be characterized by the polarization unit vector $\hat{\bf P}$, which functions as the spin component of the photon wave function. An arbitrary polarization $\hat{\bf P}$ can be represented as a superposition of two orthogonal basis vectors \cite{berestetskii1982quantum}:
\begin{eqnarray}
\hat{\bf P} = \cos(\theta_{\alpha})\hat{\bf P}_1+\sin(\theta_{\alpha})\hat{\bf P}_2\cdot e^{i\theta_{\beta}},
\end{eqnarray}
here, $\theta_{\alpha}$ denotes the angle between ${\bf P}$ and $\hat{{\bf P}}_1$, while $\theta_\beta$ represents the absolute phase. In quantum mechanics, the photon polarization state corresponding to {\bf P} can be described by the density matrix:
\begin{eqnarray}
	\rho =\frac{1}{2}\left(1+{\boldsymbol{\xi}\cdot \boldsymbol{\sigma}}\right)=\frac{1}{2}\left(\begin{matrix}
	1+\xi_{3}&\xi_{1}-i\xi_{2}\\
	\xi_{1}+i\xi_{2}&1-\xi_{3}
	\end{matrix}\right)
	\end{eqnarray}
where $\boldsymbol{\sigma}$ represents the Pauli matrix, and $\boldsymbol{\xi}=(\xi_{1},\xi_{2},\xi_{3})$ denotes the Stokes parameters, with $\xi_{1} = \sin(2\theta_\alpha)\cos(\theta_\beta)$, $\xi_{2} = \sin(2\theta_\alpha)\sin(\theta_\beta)$, $\xi_{3} = \cos(2\theta_\alpha)$.

The calculation of the probability of pair creation requires the transformation of the Stokes parameters from the initial frame of the photon ($\hat{{\bf P}}_1$, $\hat{{\bf P}}_2$, $\hat{\bf n}$) to the frame of pair production ($\hat{{\bf P}}_1'$, $\hat{{\bf P}}_2'$, $\hat{\bf n}$). The vector $\hat{{\bf P}}_1'$ is given by [${\bf E}- \hat{\bf n}\cdot(\hat{\bf n}\cdot{\bf E})+\hat{\bf n}\times{\bf B}$]/$|{\bf E}- \hat{\bf n}\cdot(\hat{\bf n}\cdot{\bf E})+\hat{\bf n}\times{\bf B}|$, and the vector $\hat{{\bf P}}_2'$ is obtained by taking the cross product of $\hat{\bf n}$ and $\hat{{\bf P}}_1'$. Here, $\hat{\bf n}$ represents the direction of propagation of the photon, and ${\bf E}$ and ${\bf B}$ denote the electric and magnetic fields, respectively. Two groups of polarization vector are connected via a rotation of angle $\psi$:
\begin{eqnarray}
	\hat{\bf P}_1^{'}&=&\hat{{\bf P}}_1 {\rm cos}(\psi)+\hat{{\bf P}}_2 {\rm sin}(\psi),\\
	\hat{\bf P}_2^{'}&=&-\hat{{\bf P}}_1 {\rm sin}(\psi)+\hat{{\bf P}}_2 {\rm cos}(\psi).
\end{eqnarray}
Thus, the Stokes parameters with respect to the vectors $\hat{\bf P}_1^{'}$, $\hat{\bf P}_2^{'}$ and $\hat{\bf n}$ are the follows:
\begin{eqnarray}
	\xi_1^{'} &=& \xi_1 {\rm cos}(2\psi)-\xi_3 {\rm sin}(2\psi),\nonumber \\
	\xi_2^{'} &=& \xi_2, \nonumber \\
	\xi_3^{'} &=& \xi_1 {\rm sin}(2\psi)+\xi_3 {\rm cos}(2\psi),
\end{eqnarray}
which is equivalent to a rotation \cite{wistisen2013vacuum,Bragin_2017}:
\begin{equation}
	\left(\begin{array}{l}
		\xi'_1 \\
		\xi'_2 \\
		\xi'_3
	\end{array}\right) =
	\left(\begin{array}{ccc}
		\cos2\psi & 0 & -\sin2\psi \\
		0 & 1 & 0 \\
		\sin2\psi & 0 & \cos2\psi
	\end{array}\right)
	\left(\begin{array}{l}
		\xi_1 \\
		\xi_2 \\
		\xi_3
	\end{array}\right) \equiv{\mathrm{ROT}}(\psi) \cdot \bm{\xi}.
	\label{rotation}
\end{equation}

\subsection{Nonlinear Breit-Wheeler pair production}
When the energy of a photon exceeds the rest mass of an electron-positron pair, i.e., $\omega_\gamma \geq 2m_ec^2$, and is subjected to an ultraintense field of $a_0 \gg 1$, the related nonlinear quantum parameter $\chi_\gamma$ can reach unity. Here, $\chi_\gamma \equiv \frac{e\hbar}{m^2c^3} \sqrt{|F^{\mu \nu}k_\nu|^2}$ and is approximately equal to $2 \omega_\gamma a_0 / m^2_e$ in the colliding geometry. In this scenario, the photon can decay into an electron-positron pair through the nonlinear Breit-Wheeler pair production (NBW) process ($\omega_\gamma + n\omega_L \rightarrow e^+ + e^-$) \cite{Bell2008}. Refs.~\cite{Chen2019,Wan_2020, Xue_2022,Li2022,Chen2022} proposed the spin- and polarization-resolved NBW MC method, and we followed the detailed methods in Ref.~\cite{Xue_2022}.
\subsubsection{NBW Probability}
The polarization-resolved NBW probability rate with dependence on the positron energy is given by
\begin{eqnarray}\label{BW}
	&&\frac{{\rm d}^2W^\pm_{\rm pair}}{{\rm d}\varepsilon_{+}{\rm d}t}= \frac{1}{2}(G_0+\xi_{1}G_1+\xi_{2}G_2+\xi_{3}G_3),
\end{eqnarray}
where the polarization-independent term $G_0$ and polarization-related terms $G_{1,2,3}$ are given by
\begin{widetext}
	\begin{eqnarray}
		\label{G0}
		G_0&=&\frac{W_{0}}{2}\left\{ {\rm IntK}_{\frac{1}{3}}(\rho)+{\frac{\varepsilon_{-}^2+\varepsilon_{+}^2}{\varepsilon_{-}\varepsilon_{+}}}{\rm K}_{\frac{2}{3}}(\rho)+\left[{\rm IntK}_{\frac{1}{3}}(\rho)-2{\rm K}_{\frac{2}{3}}(\rho)\right]
		\left({\bf S}_{-}\cdot{\bf S}_{+}\right)+\right.\nonumber\\
		&&\left.{\rm K}_{\frac{1}{3}}(\rho)\left[-\frac{\varepsilon_{\gamma}}{\varepsilon_{+}}\left({\bf S}_{+}\cdot\hat{{\bf b}}_+\right)+\frac{\varepsilon_{\gamma}}{\varepsilon_{-}}\left({\bf S}_{-}\cdot\hat{{\bf b}}_+\right)\right]+\left[\frac{\varepsilon_{-}^2+\varepsilon_{+}^2}{\varepsilon_{-}\varepsilon_{+}}{\rm IntK}_{\frac{1}{3}}(\rho)\right.\right.\nonumber\\
		&&\left.\left.-\frac{(\varepsilon_{+}-\varepsilon_{-})^2}{\varepsilon_{-}\varepsilon_{+}}{\rm K}_{\frac{2}{3}}(\rho)\right]({\bf S}_{+}\cdot\hat{\bf v}_+)({\bf S}_{-}\cdot\hat{\bf v}_+)\right\},\label{G2}\\
		G_1&=&\frac{W_{0}}{2}\left\{{\rm K}_{\frac{1}{3}}(\rho)\left[-\frac{\varepsilon_{\gamma}}{\varepsilon_{-}}({\bf S}_{+}\cdot\hat{{\bf a}}_+)+\frac{\varepsilon_{\gamma}}{\varepsilon_{+}}({\bf S}_{-}\cdot\hat{{\bf a}}_+)\right]+\frac{\varepsilon_{+}^2-\varepsilon_{-}^2}{2\varepsilon_{-}\varepsilon_{+}}{\rm K}_{\frac{2}{3}}(\rho)
		({\bf S}_{-}\times{\bf S}_{+})\cdot\hat{\bf v}_+\right.\nonumber\\
		&&\left.-\frac{\varepsilon_{\gamma}^2}{2\varepsilon_{-}\varepsilon_{+}}{\rm IntK}_{\frac{1}{3}}(\rho)\left[({\bf S}_{+}\cdot\hat{{\bf a}})({\bf S}_{-}\cdot\hat{{\bf b}})+({\bf S}_{-}\cdot\hat{{\bf a}}_+)({\bf S}_{+}\cdot\hat{{\bf b}}_+)\right]\right\},\label{G1}\\
		G_2&=&\frac{W_{0}}{2}\left\{\frac{\varepsilon_{\gamma}^2}{2\varepsilon_{-}\varepsilon_{+}}{\rm K}_{\frac{1}{3}}(\rho)({\bf S}_{-}\times{\bf S}_{+})\cdot\hat{{\bf a}}_{+}+\right.\nonumber\\
		&&\left.\frac{\varepsilon_{+}^2-\varepsilon_{-}^2}{2\varepsilon_{-}\varepsilon_{+}}{\rm K}_{\frac{1}{3}}(\rho)\left[({\bf S}_{-}\cdot\hat{\bf v}_+)
		({\bm S}_{+}\cdot\hat{{\bf b}}_+)+({\bf S}_{+}\cdot\hat{\bf v}_+)({\bf S}_{-}\cdot\hat{{\bf b}}_+)\right]+\right.\nonumber\\&&\left.\left[\frac{\varepsilon_{\gamma}}{\varepsilon_{-}}{\rm IntK}_{\frac{1}{3}}(\rho)-\frac{\varepsilon_{+}^2-\varepsilon_{-}^2}{\varepsilon_{-}\varepsilon_{+}}{\rm K}_{\frac{2}{3}}(\rho)\right]({\bf S}_{-}\cdot\hat{\bf v}_+)+\right.\nonumber\\&&\left.\left[\frac{\varepsilon_{\gamma}}{\varepsilon_{+}}{\rm IntK}_{\frac{1}{3}}(\rho)+\frac{\varepsilon_{+}^2-\varepsilon_{-}^2}{\varepsilon_{-}\varepsilon_{+}}{\rm K}_{\frac{2}{3}}(\rho)\right]({\bf S}_{+}\cdot\hat{\bf v}_+)  \right\}, \\
		G_3&=&\frac{W_{0}}{2}\left\{-{\rm K}_{\frac{2}{3}}(\rho)+\frac{\varepsilon_{-}^2+\varepsilon_{+}^2}{2\varepsilon_{-}\varepsilon_{+}}{\rm K}_{\frac{2}{3}}(\rho)({\bf S}_{-}\cdot{\bf S}_{+})-\right.\nonumber\\&&\left.{\rm K}_{\frac{1}{3}}(\rho)\left[\frac{\varepsilon_{\gamma}}{\varepsilon_{+}}({\bf S}_{-}\cdot\hat{{\bf b}}_+)-\frac{\varepsilon_{\gamma}}{\varepsilon_{-}}({\bf S}_{+}\cdot\hat{{\bf b}}_+)\right]+\right.\nonumber\\&&\left.\frac{\varepsilon_{\gamma}^2}{2\varepsilon_{-}\varepsilon_{+}}{\rm IntK}_{\frac{1}{3}}(\rho)\bigg [({\bf S}_{+}\cdot\hat{{\bf b}}_+)({\bf S}_{-}\cdot\hat{{\bf b}}_+)\right.-\nonumber\\
		&&\left.({\bf S}_{+}\cdot\hat{{\bf a}}_+)({\bf S}_{-}\cdot\hat{{\bf a}}_+)\bigg ]-\frac{(\varepsilon_{+}-\varepsilon_{-})^2}{2\varepsilon_{-}\varepsilon_{+}}{\rm K}_{\frac{2}{3}}(\rho)({\bf S}_{+}\cdot\hat{\bf v}_+)({\bf S}_{-}\cdot\hat{\bf v}_+)\right\},\label{G3}
	\end{eqnarray}
\end{widetext}
where $W_{0} = \alpha / \left(\sqrt{3}\pi \omega'^2_\gamma \right)$, $\omega'_\gamma = \varepsilon_\gamma / m_ec^2$, $\rho = 2\varepsilon_{\gamma}^2/\left(3\chi_{\gamma}\varepsilon_{-}\varepsilon_{+}\right) = 2/\left[3\delta(1-\delta)\right]$, $\delta = \varepsilon_+ / \varepsilon_\gamma$, ${\rm IntK}_{\frac{1}{3}}(\rho)\equiv \int_{\rho}^{\infty} {\rm d}z {\rm K}_{\frac{1}{3}}(z)$,  ${\rm K}_n$ is the $n$-order modified Bessel function of the second kind, $\alpha$ the fine structure constant, $\varepsilon_{\gamma}$, $\varepsilon_{-}$ and $\varepsilon_{+}$ the energies of parent photon, created electron and positron, respectively, $\hat{\bf v}_+ = {\bf v}_+/|{\bf v}_+|$ with  the positron velocity ${\bf v}_+$, $\hat{{\bf a}}_+={\bf a}_+/|{\bf a}_+|$ with  the positron acceleration ${\bf a}_+$ in the rest frame of positron, $\hat{{\bf b}}_+={\bf v}_+\times{\bf a}_+/|{\bf v}_+\times{\bf a}_+|$, $\xi_{1}$, $\xi_{2}$ and $\xi_{3}$ are the Stokes parameters of $\gamma$ photon, and ${\bf S}_{+}$ (${\bf S}_{-}$) denotes the positron (electron) spin vector. Note that the Stokes parameters must be transformed from the photon initial frame ($\hat{{\bf P}}_1$, $\hat{{\bf P}}_2$, $\hat{\bf n}$) to the pair production frame ($\hat{{\bf P}}_1'$, $\hat{{\bf P}}_2'$, $\hat{\bf n}$); see transformations of the  Stokes parameters  in Sec.~\ref{StokesTra.}.

Summing over the electron spin, the pair production depending on the positron spin ${\bf S}_+$ and the photon polarization $\boldsymbol{\xi}$ is obtained:

\begin{eqnarray}\label{BW_pos}
	&&\frac{{\rm d}^2W_{\rm pair}^{+}}{{\rm d}\varepsilon_{+}{\rm d}t}= W_{0}\bigg \{ {\rm IntK}_{\frac{1}{3}}(\rho)+\frac{\varepsilon_{-}^2+\varepsilon_{+}^2}{\varepsilon_{-}\varepsilon_{+}}{\rm K}_{\frac{2}{3}}(\rho)-\frac{\varepsilon_{\gamma}}{\varepsilon_{+}}{\rm K}_{\frac{1}{3}}(\rho)({\bf S}_{+}\cdot\hat{{\bf b}}_+)\nonumber\\
	&&-\xi_{1}\bigg[\frac{\varepsilon_{\gamma}}{\varepsilon_{-}}{\rm K}_{\frac{1}{3}}(\rho)({\bf S}_{+}\cdot\hat{{\bf a}}_+)\bigg]+\xi_{2}\left[\frac{\varepsilon_{+}^2-\varepsilon_{-}^2}{\varepsilon_{-}\varepsilon_{+}}{\rm K}_{\frac{2}{3}}(\rho)+\frac{\varepsilon_{\gamma}}{\varepsilon_{+}}{\rm IntK}_{\frac{1}{3}}(\rho)\right] \times \nonumber\\
	&&\left.({\bf S}_{+}\cdot \hat{\bf v}_{+})\right]-\xi_{3}\left[{\rm K}_{\frac{2}{3}}(\rho)-\frac{\varepsilon_{\gamma}}{\varepsilon_{-}}{\rm K}_{\frac{1}{3}}(\rho)({\bf S}_{+}\cdot\hat{{\bf b}}_+)\right]\bigg \}.
\end{eqnarray}
It can be rewritten as:
\begin{eqnarray}
	\label{BWposAxis}
	\frac{{\rm d}^2W_{\rm pair}^{+}}{{\rm d}\varepsilon_{+}{\rm d}t} &=& W_{0}(C+{\bf S}_{+}\cdot{\bf D}),
\end{eqnarray}
where
\begin{eqnarray}
	\label{BWposAxis1}
	C &=& {\rm IntK}_{\frac{1}{3}}(\rho)+\frac{\varepsilon_{-}^2+\varepsilon_{+}^2}{\varepsilon_{-}\varepsilon_{+}}{\rm K}_{\frac{2}{3}}(\rho)-\xi_{3}{\rm K}_{\frac{2}{3}}(\rho),\\
	{\bf D}&=&-\left(\frac{\varepsilon_{\gamma}}{\varepsilon_{+}}-\xi_{3}\frac{\varepsilon_{\gamma}}{\varepsilon_{-}}\right){\rm K}_{\frac{1}{3}}(\rho)\hat{{\bf b}}_+-\xi_{1}\frac{\varepsilon_{\gamma}}{\varepsilon_{-}}{\rm K}_{\frac{1}{3}}(\rho)\hat{{\bf a}}_++\nonumber\\
	&&\xi_{2}\bigg[ \frac{\varepsilon_{+}^2-\varepsilon_{-}^2}{\varepsilon_{-}\varepsilon_{+}}{\rm K}_{\frac{2}{3}}(\rho)+\frac{\varepsilon_{\gamma}}{\varepsilon_{+}}{\rm IntK}_{\frac{1}{3}}(\rho)\bigg]\hat{\bf v}_+.\label{BWposAxis2}
\end{eqnarray}
When a photon decays to pair, the positron spin state is instantaneously collapsed into one of its basis states defined by the instantaneous SQA, along the energy-resolved average polarization ${\bf S}_+^{(\varepsilon_{+})} = {\bf D} /C$.

Similarly, summing over the positron spin, the pair production probability depending on the electron spin ${\bf S}_-$ and the photon polarization is obtained:
\begin{eqnarray}
	\label{BWeleAxis}
	&&\frac{{\rm d}^2W_{\rm pair}^{-}}{{\rm d}\varepsilon_{+}{\rm d}t}= W_{0}(C+{\bf S}_{-}\cdot{\bf D}^{'}),\\
	&&{\bf D}^{'}=\left(\frac{\varepsilon_{\gamma}}{\varepsilon_{-}}-\xi_{3}\frac{\varepsilon_{\gamma}}{\varepsilon_{+}}\right){\rm K}_{\frac{1}{3}}(\rho)\hat{{\bf b}}_++\xi_{1}\frac{\varepsilon_{\gamma}}{\varepsilon_{+}}{\rm K}_{\frac{1}{3}}(\rho)\hat{{\bf a}}_+\nonumber\\
	&&-\xi_{2}\bigg[ \frac{\varepsilon_{+}^2-\varepsilon_{-}^2}{\varepsilon_{-}\varepsilon_{+}}{\rm K}_{\frac{2}{3}}(\rho)-\frac{\varepsilon_{\gamma}}{\varepsilon_{-}}{\rm IntK}_{\frac{1}{3}}(\rho)\bigg]\hat{\bf v}_+.\label{BWeleAxis1}
\end{eqnarray}

The pair production probability, relying solely on photon polarization, is determined by summing over both positron and electron spins:
\begin{equation}
 \label{BWprob}
 \frac{{\rm d}^2W_{\rm pair}}{{\rm d}\varepsilon_+ {\rm d}t}=2W_0\left\{{\rm IntK}_{\frac{1}{3}}(\rho)+\frac{\varepsilon_-^2+\varepsilon_+^2}{\varepsilon_-\varepsilon_+}{\rm K}_{\frac{2}{3}}(\rho)-\xi_{3}{\rm K}_{\frac{2}{3}}(\rho)\right\}.
\end{equation}

\subsubsection{MC algorithm}
\begin{figure}[ht]
	\centering
	\includegraphics[width=\linewidth]{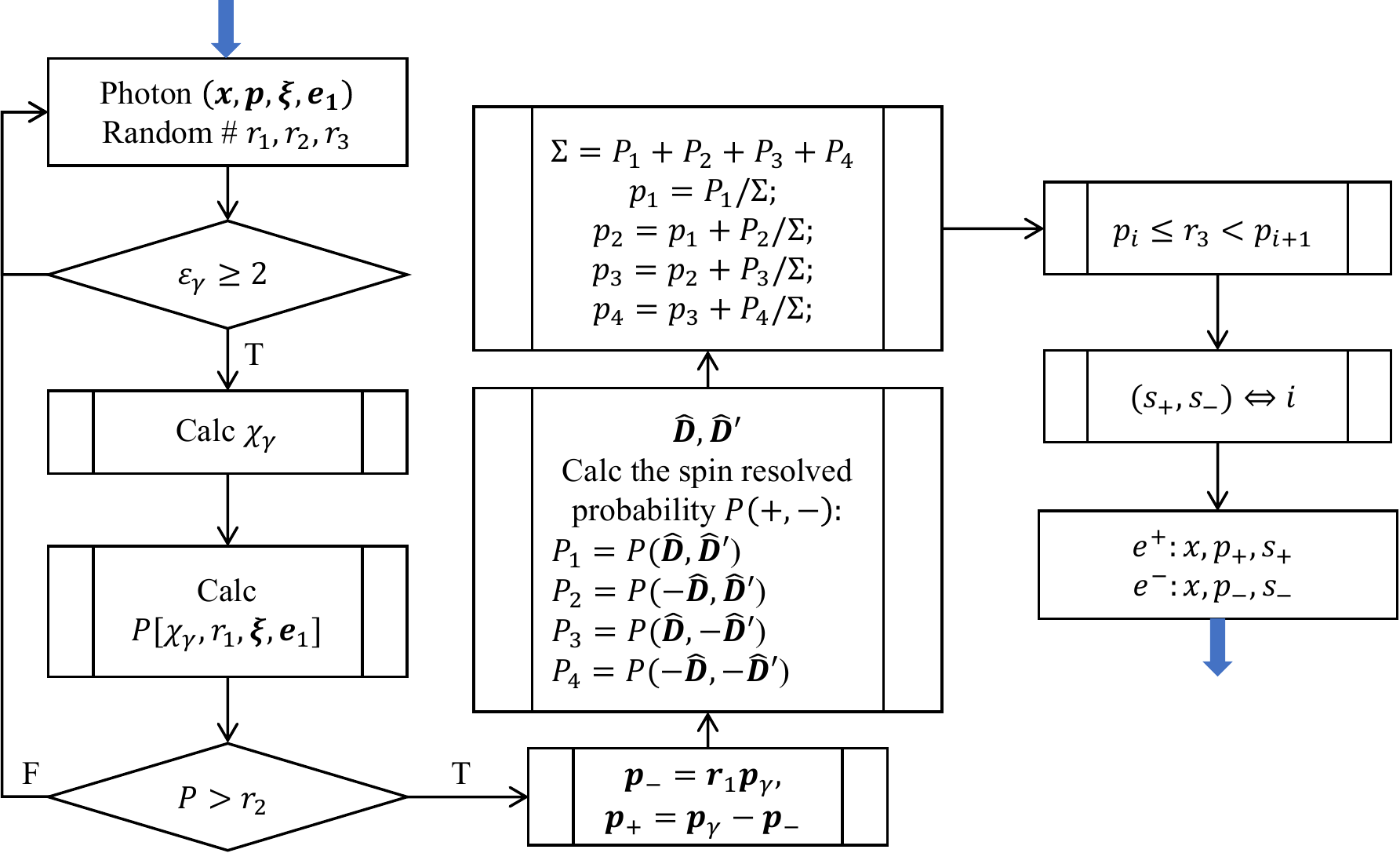}
	\caption{Flowchart of the spin- and polarization-resolved nonlinear Breit-Wheeler (NBW) pair production process.}
	\label{nbw-loop}
\end{figure}
\begin{figure}[ht]
	\centering
	\includegraphics[width=\linewidth]{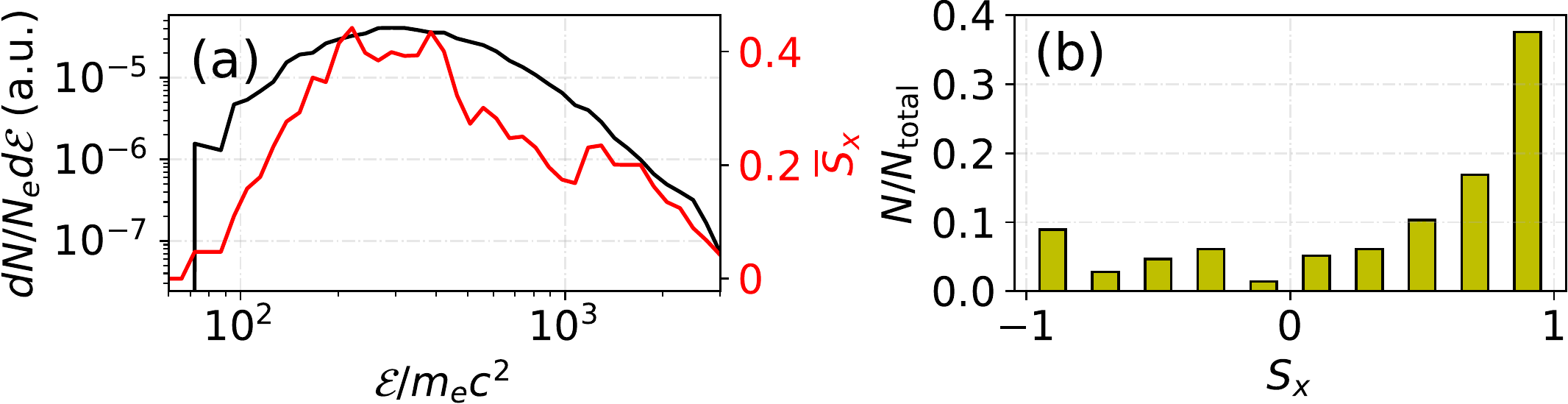}
	\caption{(a) Normalized energy spectra (black solid line) and energy-resolved longitudinal spin polarization (red solid line) of positrons. (b) Statistics of the longitudinally spin components of generated positrons. The laser and electron beam parameters are consistent with those in Figure.~\ref{fig-beam-dynamics}.} \label{fig-nbw}
\end{figure}

The algorithm for simulating pair creation with polarization is illustrated in Figure.~\ref{nbw-loop}. At every simulation step $\Delta t$, a pair is generated with positron energy $\varepsilon_+ = r_1 \varepsilon_\gamma$ when the probability density $P \equiv {\rm d^2}W_{\rm pair}/{\rm d}\varepsilon_+{\rm d}t \cdot \Delta t$ of pair production is greater than or equal to a random number $r_2$ within the range [0,1). Here, ${\rm d^2}W_{\rm pair}/{\rm d}\varepsilon_+{\rm d}t $ is computed using Equation (\ref{BWprob}). The momentum of the created positron (electron) is parallel to that of the parent photon, and the energy of the electron $\varepsilon_{-}$ is determined as $\varepsilon_{\gamma}-\varepsilon_{+}$. The final spin states of the electron and positron are determined by the four probability densities $P_{1,2,3,4}$, each representing spin parallel or antiparallel to SQA, where $P_{1,2,3,4}$ is computed from Equation (\ref{BW}). Finally, a random number $r_3$ is used to sample the final spin states for the electron and positron. Note that, here all random numbers are sampled uniformly from $[0, 1)$ as in the NCS algorithm.
An example of the production of secondary electrons and positrons resulting from a collision between a laser and an electron beam is illustrated in Figure.~\ref{fig-nbw}.

\subsection{High-energy bremsstrahlung}

The high energy bremsstrahlung is another important emission mechanism, which can also be modeled using a MC collision model \cite{wan_epjd}. The MC collision model was tested using the Geant4 code \cite{Agostinelli_2003}, and the results are presented in the following section. The bremsstrahlung emission is described as described by the cross-section in Ref.~\cite{Tsai1974}
\begin{equation}
	\begin{aligned}
		\frac{d\sigma_{eZ}}{d\omega}(\omega, y) = & \frac{\alpha r_0^2}{\omega} \lbrace (\frac{4}{3} - \frac{4}{3}y + y^2) \\  & \times [ Z^2(\phi_1 - \frac{4}{3} \mathrm{ln}Z - 4f) + Z(\psi_1 - \frac{8}{3}\mathrm{ln}Z)] \\ & +
		\frac{2}{3}(1-y)[Z^2(\phi_1 - \phi_2) + Z(\psi_1 - \psi_2)] \rbrace,
	\end{aligned}
\end{equation}
where $y = \hbar \omega / E_e$ is the energy ratio of the emitted photon to the incident electron, $r_0$ is the classical electron radius, functions $\phi_{1,2}$ and $\psi_{1,2}$ depend on the screening potential by atomic electrons, while the Coulomb correction term is denoted by $f$. When the atomic number of the target is greater than $5$, we use Eqs.~(3.38-3.41) from Ref.~\cite{Tsai1974} to calculate these functions. However, for targets with $Z<5$, the approximated screen functions are unsuitable and require modification.

The PENELOPE code \cite{PENELOPE} utilizes another method that involves tabulated data from Ref.~\cite{Seltzer_1986}. This method transforms the ``scaled" bremsstrahlung differential cross-section (DCS) to differential cross-section by using the following equation \cite{PENELOPE}:
\begin{equation}
\frac{d\sigma_{br}}{d \omega} = \frac{Z^2}{\beta^2} \frac{1}{\omega}\chi(Z, E_e, y),
\end{equation}
where $\beta = v/c$ is the normalized electron velocity. Integrating this expression over the photon frequencies yields a tabulated total cross-section $\sigma_{br}(E_e, y)$ MC simulation, i.e, the direct sampling method can be used.

The DCS for electron and positron are related by the equation
\begin{equation}
\frac{d \sigma_{br}^{+}}{d\omega} = F_p(Z, E_e) \frac{d \sigma_{br}^{-}}{d\omega},
\end{equation}
where $F_p(Z, E_e)$ is an analytical approximation factor that can be found in Ref.~\cite{PENELOPE}. This reference demonstrates a high level of accuracy, with a difference of only approximately $0.5\%$ compared to Ref.~\cite{Kim1986}.

The bremsstrahlung implementation is based on a direct MC sampling. Given an incident electron with energy $E_e$ and velocity $v$, the probability of triggering a bremsstrahlung event is calculated as $P_{br} = 1 - e^{\Delta s/ \lambda}$, where $\Delta s = v \Delta t$, $v = |\vec{v}|$ is the incident particle velocity, $\Delta t$ is the time interval, $\lambda = 1 / n\sigma(E_e)$, $n$ represents the target particle density and $\sigma(E_e)$ is the total cross-section, respectively. A random number $r_1$ is then generated and compared to $P_{br}$. If $r_1 < P_{br}$, a bremsstrahlung event is triggered. The energy of the resulting photon is determined by generating another random number $r_2$, which is then multiplied by $\sigma_{br}(E_e)$ to obtain the energy ratio $y$ through $\sigma(y, E_e) = \sigma(E_e) r_2$. Finally, a photon with energy $\hbar \omega = E_e y$ and momentum direction $\vec{k}/|\vec{k}| = \mathbf{v} / |\mathbf{v}|$ is generated. To improve computational efficiency, low energy photons are discarded by setting a minimum energy threshold. This probabilistic approach is similar to the method used to calculate the random free path \cite{PENELOPE}. The implementation of Bethe-Heitler pair production follows a similar process.

\begin{figure}[ht]
	\centering
	\includegraphics[width=\linewidth]{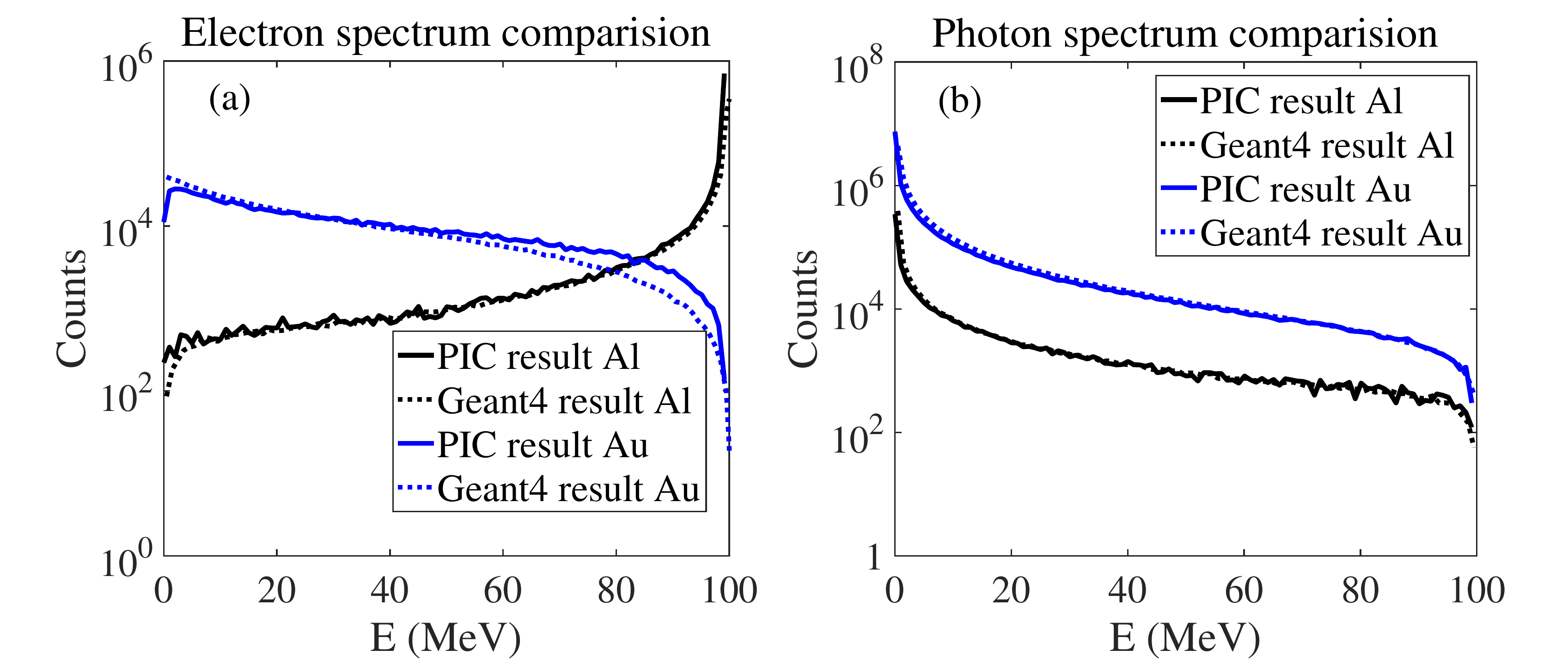}
	\caption{(color online). Bremsstrahlung of 100 MeV electrons. (a) for the scattered electron spectra, and (b) for the yield photon spectra. Solid lines represent PIC results and dashed lines represent Geant4 results. These figures are obtained from Ref. \cite{wan_epjd}}
	\label{mevbrems}
\end{figure}

\begin{figure}[ht]
	\centering
	\includegraphics[width=\linewidth]{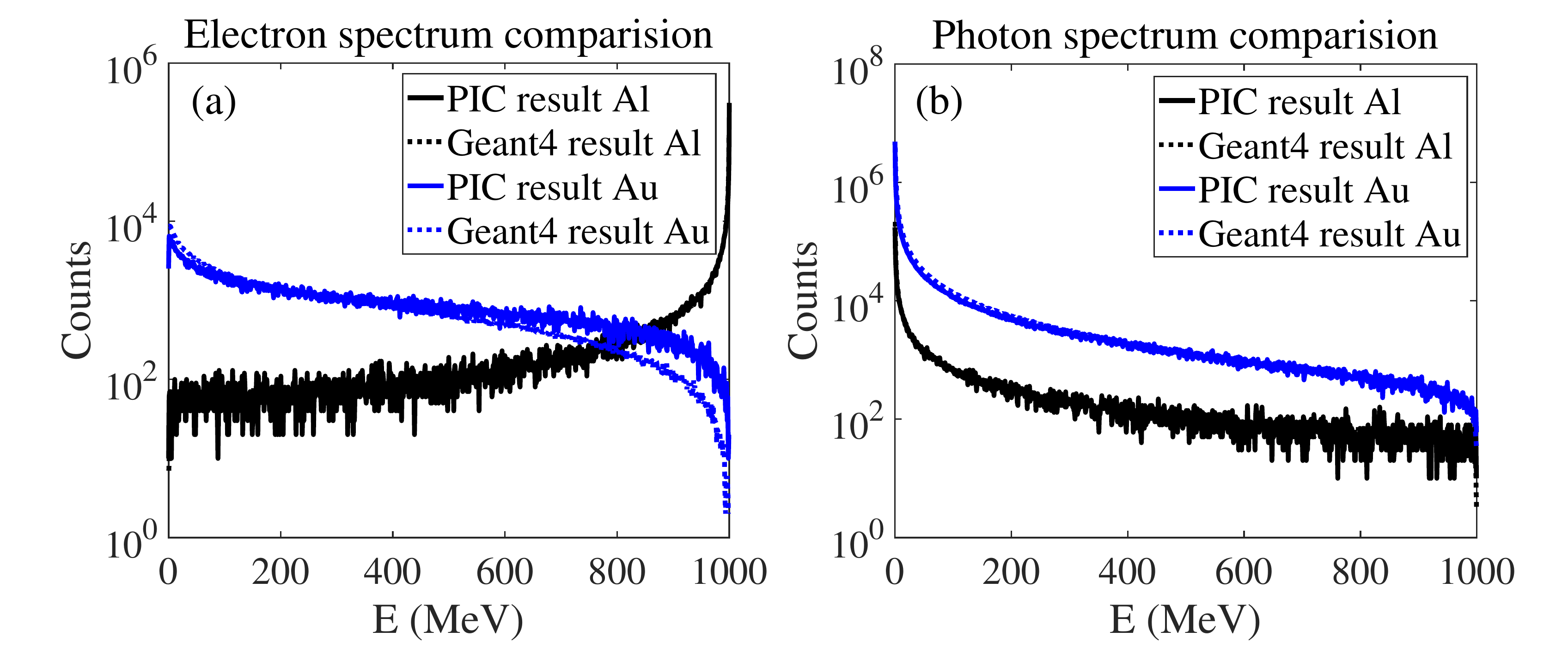}
	\caption{(color online). Bremsstrahlung of 1 GeV electrons. (a) for the scattered electron spectra, and (b) for the yield photon spectra. Solid lines represent PIC results and dashed lines represent Geant4 results. These figures are obtained from Ref. \cite{wan_epjd}.}
	\label{gevbrems}
\end{figure}

The implementation of Bremsstrahlung emission was tested using the Geant4 software \cite{Agostinelli_2003}, which is widely used for modeling high-energy particle scattering with detectors. In this study, we utilized electron bunches of 1 GeV and 100 MeV with $10^5$ primaries to collide with a 5 mm Au target with $Z=79$, $\rho=19.3$ $\mathrm{g/cm^3}$ and a 5 mm Al target with $Z=13$, $\rho=2.7$ $\mathrm{g/cm^3}$. We disabled the field updater and weighting procedure in the PIC code, and only enabled the particle pusher and Bremsstrahlung MC module. The electron and photon spectra were found to be in good agreement with the Geant4 results, except for a slightly higher photon emission in the high energy tail (which is due to the difference in the cross-section data). Figure.~\ref{mevbrems} displays the spectra of electrons and photons from a 100 $\mathrm{MeV}$ electron bunch normally incident onto the aluminum and gold slabs, while similar distributions for a 1 $\mathrm{GeV}$ electron bunch are shown in Figure.~\ref{gevbrems}.

\subsection{Vacuum birefringence}
Another important process for polarized photons in ultraintense laser matter interactions is vacuum birefringence (VB), in addition to the NBW processes. In this paper, we utilize Eq.(4.26) from Ref.~\cite{Shore_2007_Superluminality} to calculate the refractive index $n$ for a photon with arbitrary energy $\omega$ (wavelength $\lambda$) in a constant weak EM field [$|E|(|B|) \ll E_{cr}$]. We include the electric field and assume relativistic units $c=\hbar=1$. The resulting expression is:
\begin{equation}
	n \approx 1 - \frac{\alpha\chi_\gamma^2m^2}{16\pi\omega^2} \int_{-1}^1 d\upsilon (1-\upsilon^2) \left\{ \begin{matrix} \frac{1}{2}(1 + \frac{1}{3}\upsilon^2) \\ 1 - \frac{1}{3}\upsilon^2 \end{matrix} \right\} \left[ \pi x^{4/3} \mathrm{Gi}'(x^{2/3}) - i \frac{x^2}{\sqrt{3}}\mathrm{K}_{2/3} \left( \frac{2}{3}x \right) \right], \label{eqnn}
	\end{equation}
where $\alpha$ is the fine structure constant, $m$ is the electron mass, $\chi_\gamma$ is the nonlinear quantum parameter as defined before, $x = \frac{4}{(1-\upsilon^2)\chi_\gamma}$, ${\rm Gi}'(x)$ is the derivative of the Scorer's function, and ${\rm K}_n(x)$ is the $n$th-order modified Bessel function of the second kind~[38]. ${\bf E}_{\rm red,\perp} = {\bf E}_\perp + \hat{k} \times {\bf B}_\perp$ is the transverse reduced field (acceleration field for electrons). The first and second columns correspond to the eigenmodes parallel and perpendicular to the reduced field, respectively.
By extracting a factor of
\[
\mathcal{D} = \frac{\alpha}{90\pi} \left(\frac{e|{\bf E}_{\rm red\perp}|}{m^2}\right)^2 \equiv \frac{\alpha}{90\pi} \frac{\chi_\gamma^2}{\omega^2/m^2}.
\]
Eq.~(\ref{eqnn}) arrives at
\begin{eqnarray}
	\mathrm{Re}(n) &=&1 - \frac{45}{4} \mathcal{D} \int_0^1 d\upsilon  (1-\upsilon^2) \left\{ \begin{matrix} \frac{1}{2}(1 + \frac{1}{3}\upsilon^2) \\ 1 - \frac{1}{3}\upsilon^2 \end{matrix}  \right\} \left[ \pi x^{4/3} \mathrm{Gi}'(x^{2/3}) \right], \\
	\mathrm{Im}(n) &=& \frac{45}{4} \mathcal{D} \int_0^1 d\upsilon  (1-\upsilon^2) \left\{ \begin{matrix} \frac{1}{2}(1 + \frac{1}{3}\upsilon^2) \\ 1 - \frac{1}{3}\upsilon^2 \end{matrix}  \right\} \left[\frac{x^2}{\sqrt{3}}\mathrm{K}_{2/3} \left( \frac{2}{3}x \right) \right].
\end{eqnarray}

In the weak field limit of $\chi_\gamma \ll 1$, the imaginary part associated with pair production, is negligible. Now we define
\begin{equation}
	\begin{array}{l}
		M(\chi_\gamma) = - \frac{45}{4} \int_0^1 d\upsilon  (1-\upsilon^2) \left\{ \begin{matrix} \frac{1}{2}(1 + \frac{1}{3}\upsilon^2) \\ 1 - \frac{1}{3}\upsilon^2 \end{matrix}  \right\} \left[ \pi x^{4/3} \mathrm{Gi}'(x^{2/3}) \right], ~\mathrm{yielding} \\
		\mathrm{Re}(n) = 1 + M(\chi_\gamma) \mathcal{D} \equiv 1 + M(\chi_\gamma) \frac{\alpha}{90\pi} \frac{\chi_\gamma^2}{\omega^2/m^2}.
	\end{array}
\end{equation}

\begin{figure}[ht]
	\begin{center}
	\includegraphics[width=0.65\linewidth]{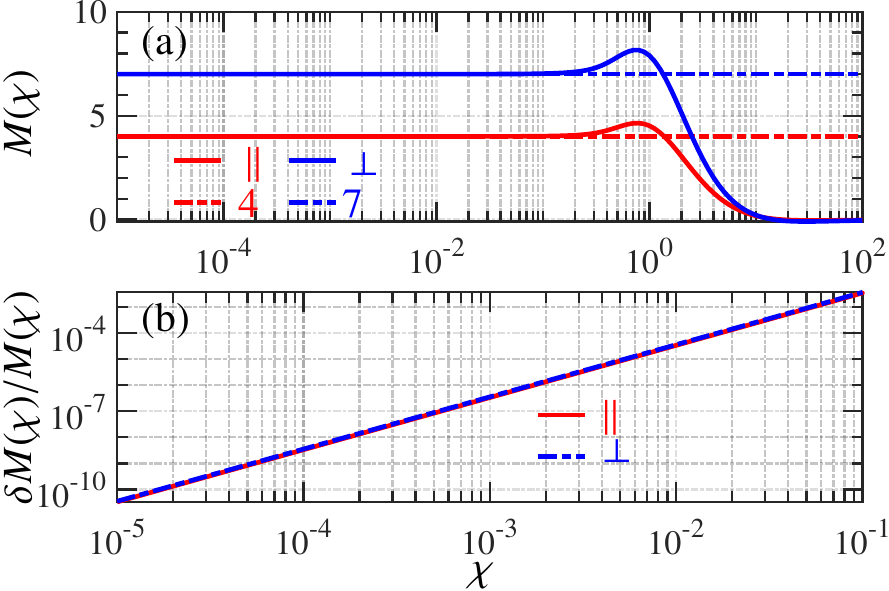}
	\caption{(a): $M(\chi_\gamma)$ and the corresponding low-energy-limit constant with red and blue dash-dotted lines equal to 4 and 7, respectively. (b): Relative error between $M(\chi_\gamma)$ and the low-energy-limit constant.} \label{fig1}
\end{center}
\end{figure}

The numerical results of $M(\chi_\gamma)$ and comparisons with the low-energy-limit ($\omega_\gamma \ll m$) constants are given in Figure.~\ref{fig1}.
While, in the limit of $\chi_\gamma \ll 1$, the real part simplifies to
\begin{eqnarray}
	\mathrm{Re}(n)=1+ \mathcal{D}
	\left\{ \begin{matrix}
		4_+ \\ 7_-
	\end{matrix} \right\}, \label{eqn26}
\end{eqnarray}
and can be used to simulate the vacuum birefringence (VB) effect with good accuracy for $\chi_\gamma \ll 1$. Note that these results are identical to those in Refs.~\cite{Shore_2007_Superluminality,Karbstein_2013_Photon,Dinu2014}.
For large $\chi_{\gamma}$, two interpolated refractive indexes are used.

The phase retardation between two orthogonal components is given by $\delta\phi = \phi_+ - \phi_- = \Delta n \frac{2\pi l}{\lambda} = -3\mathcal{D} \frac{2\pi l}{\lambda}$ with $l$ denotes the propagation length, and the VB effect is equivalent to a rotation of the Stokes parameters:
\begin{equation}
	\left(\begin{array}{l}
		\xi'_1 \\
		\xi'_2 \\
		\xi'_3
	\end{array}\right) =
	\left(\begin{array}{ccc}
		\cos\delta\phi & -\sin\delta\phi & 0 \\
		\sin\delta\phi & \cos\delta\phi & 0 \\
		0 & 0 & 1
	\end{array}\right)
	\left(\begin{array}{l}
		\xi_1 \\
		\xi_2 \\
		\xi_3
	\end{array}\right) \equiv{\mathrm{QED}}(\delta\phi) \cdot \boldsymbol{\xi}.
	\label{rotation2}
\end{equation}

\begin{figure}[ht]
	\centering
	\includegraphics[width=\linewidth]{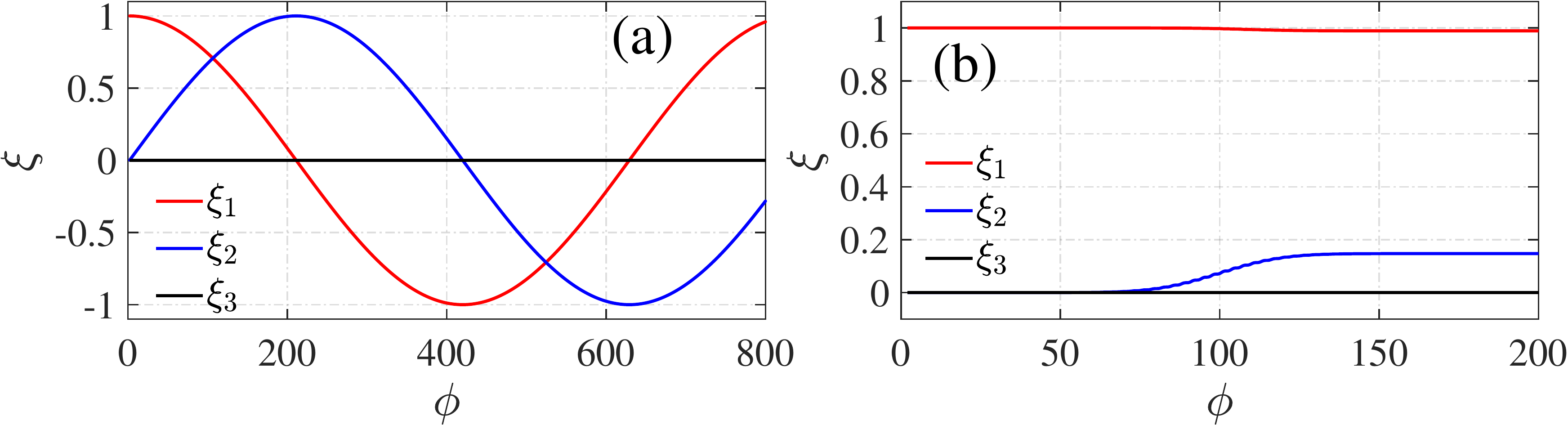}
	\caption{VB effect of a $\gamma$-photon [$\varepsilon_\gamma = 1$ GeV, $\xi = (1, 0, 0)$] propagating through a (a) static crossed field with $E_y = -B_z = 100$ and (b) laser field (same as those in Figure.~\ref{fig-spin}).}\label{fig-vbeffect}
\end{figure}

The VB effect of the probe photons in the Particle-In-Cell code is simulated with the following algorithms \cite{wan2022enhanced}:

\begin{algorithm}[H]
	\SetAlgoLined
	\emph{Initialization part}\;
	\textbf{PIC initialization}\;
	\ForEach{photon in photonList}{
		photon.$\boldsymbol{\xi}$ = $\boldsymbol{\xi}_0$\;
		photon.$\hat{a}_\pm$ = ($\hat{x},\hat{y}$)\;
	}
	\emph{evolution part}\;
	\While{not final step}{
		\textbf{do PIC loop ...}\;
		\ForEach{photon}{
			get $\mathbf{E},\mathbf{B}$\;
			get $\theta$, and $\hat{a}_+(\theta) \parallel \mathbf{E}_{\rm red\perp},\hat{a}_-(\theta) = \hat{k}\times\hat{a}(\theta)_+$\;
			rotate $\boldsymbol{\xi}$ from $\hat{a}_\pm$ to $\hat{a}_\pm(\theta)$ via Eqs. (\ref{rotation})\;
			calculate new $\boldsymbol{\xi}$ via Eq. (\ref{rotation2}) \;
			update the polarization basis: photon.$\hat{a}_\pm = \hat{a}_\pm(\theta)$.
		}
	}
	\emph{Post-processing}\;
	\emph{select a detection plane (polarization basis), for instance $\hat{x},\hat{y}$}\;
	\ForEach{$x,y$ in the detector}{
		\ForEach{photon in this area}{
			rotate $\boldsymbol{\xi}$ from $\hat{a}_\pm$ to $(\hat{x},\hat{y})$ via Eqs. (\ref{rotation})\;
			average all $\boldsymbol{\xi}$.
		}
	}
	\caption{VB effect in SLIPs}
\end{algorithm}
See an example of VB effect in Figure.~\ref{fig-vbeffect}.

\section{Framework of SLIPs}
\begin{figure}[ht]
	\begin{center}
		\includegraphics[width=\linewidth]{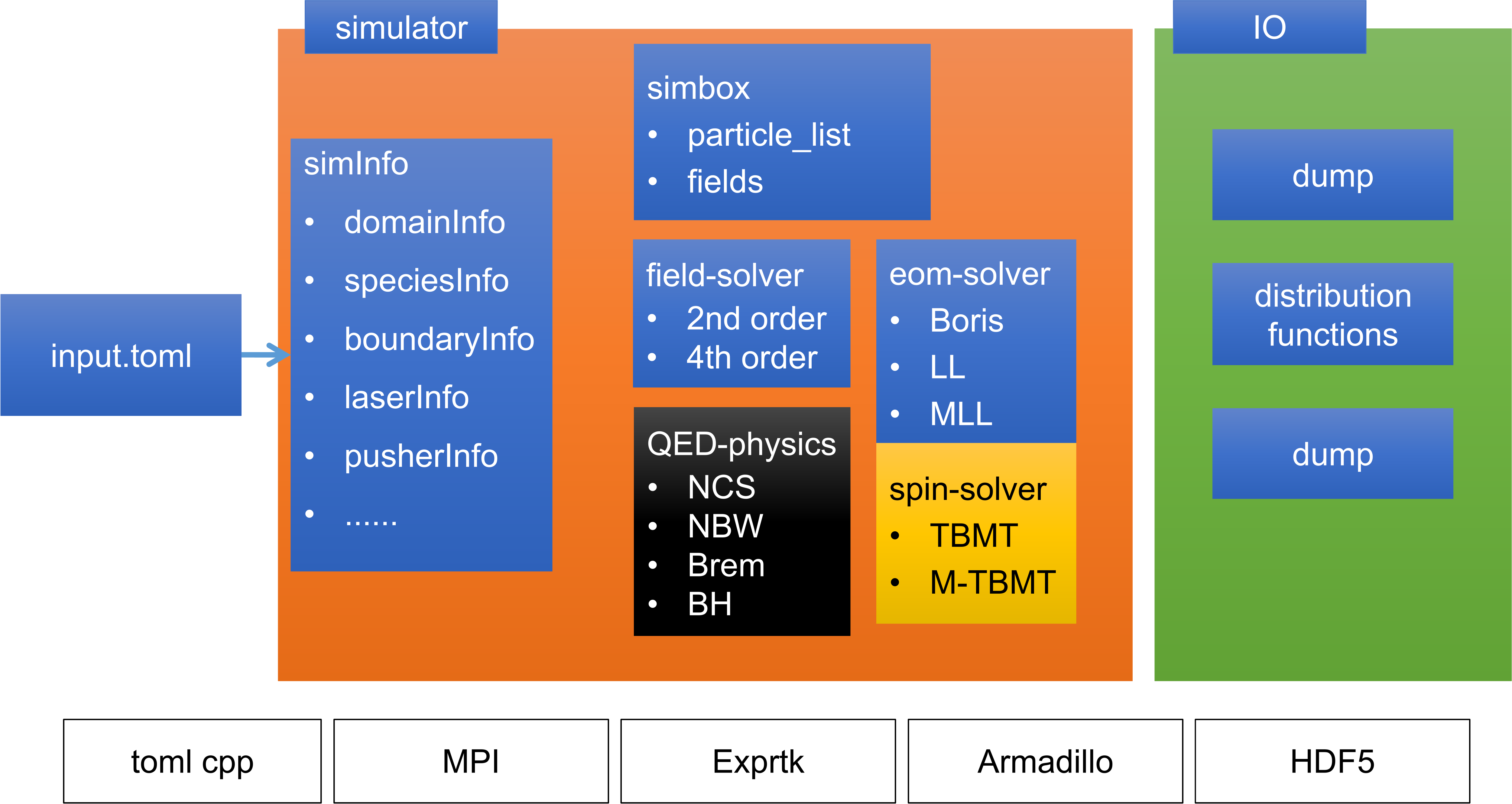}
		\caption{Data structure of the SLIPs.}\label{datastructure}
	\end{center}
\end{figure}

\begin{figure}[ht]
	\begin{center}
		\includegraphics[width=\linewidth]{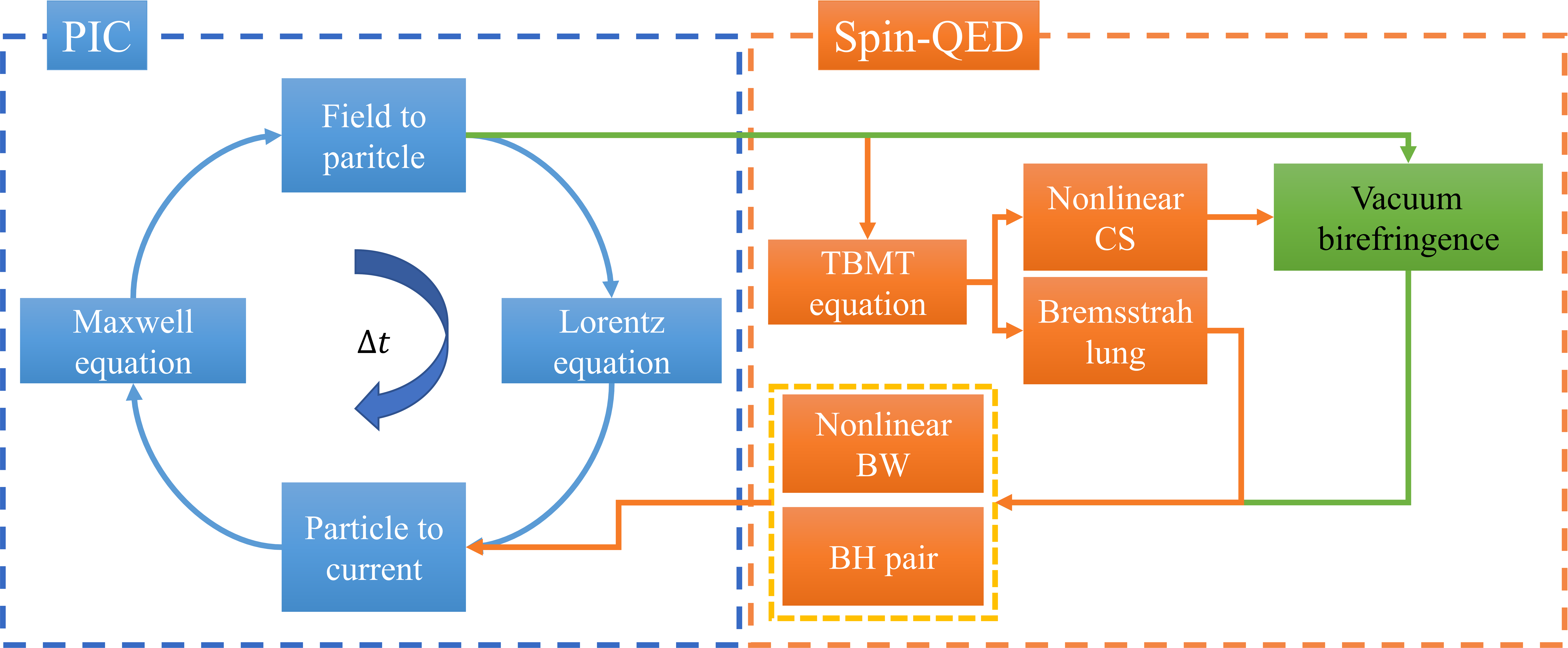}
		\caption{Framework of the SLIPs.}\label{layout}
	\end{center}
\end{figure}

These physical processes have been incorporated into the spin-resolved laser-plasma interaction simulation code, known as SLIPs. The data structure and framework layout are illustrated in Figs.~\ref{datastructure} and \ref{layout}. 

As depicted in Figure.~\ref{datastructure}, SLIPs utilize a \texttt{toml} file to store simulation information, which is then parsed into a \texttt{SimInfo} structure that includes \texttt{domainInfo}, \texttt{speciesInfo}, \texttt{boundaryInfo}, \texttt{laserInfo}, \texttt{pusherInfo}, and other metadata. 
Subsequently, this metadata is employed to generate a \texttt{SimBox} that comprises all \texttt{ParticleList} and \texttt{Fields}, and define the \texttt{FieldSolver}, \texttt{EOMSolver} and initialize QED processes.

The internal data structure of SLIPs is constructed using the open-source numerical library, Armadillo C++ \cite{Sanderson_2016,Sanderson_2018}. String expressions are parsed using the Exprtk library \cite{exprtk}. The data is then dumped using serial-hdf5 and merged with external Python scripts to remove ghost cells.

The spin-resolved processes, i.e., tagged as Spin-QED in the diagram in Figure.~\ref{layout}, are implemented in conjunction with the Lorentz equation. In the coding, the Spin-QED part is arranged as a sequential series of processes. For example, Lorentz and T-BMT are followed by radiative correction, VB, NBW, and NCS with Bremsstrahlung: $\texttt{Lorentz \& T-BMT} \Rightarrow \texttt{Radiative correction} \Rightarrow \texttt{VB} \Rightarrow \texttt{NBW} \Rightarrow \texttt{NCS \& Bremss}$.

\section{Polarized particles simulations}
In this section, we present known results that were calculated from the single-particle mode using the SLIPs. The spin-resolved NCS/NBW are evaluated by generating spin-polarized electrons/positrons. The simulation setups used in this study are identical to those described in Refs.~\cite{Li2019} and \cite{Wan_2020}.

\subsection{Polarized electron/positron simulation}
\begin{figure}[ht]
	\begin{center}
		\includegraphics[width=\linewidth]{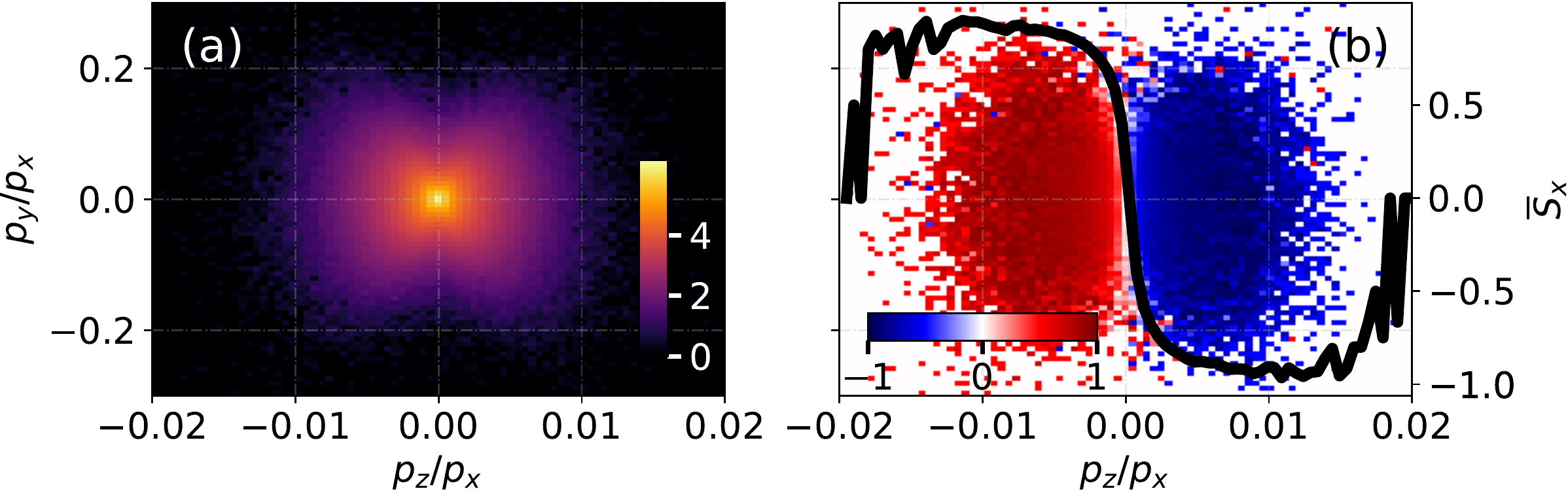}
		\caption{Generation of polarized electrons: (a) number density $dN/d\theta_xd\theta_y$, (b) spin polarization $S_x$.}
	\end{center}\label{fig-electron-pol}
\end{figure}
To simulate the generation of spin-polarized electrons, we utilized an elliptically polarized laser with an intensity of $a_0 = 30$, a wavelength of $\lambda_0 = 1\mathrm{\mu m}$, and an ellipticity of $a_{y,0} / a_{x, 0} = 3\%$. This laser was directed towards an ultrarelativistic electron bunch with an energy of 10 GeV, which was produced through laser-wakefield acceleration. The resulting polarized electrons are depicted in Figure.~\ref{fig-electron-pol}, and shows good agreement with the previously published results in Ref.~\cite{Li2022}.

\subsection{Polarized $\gamma$-photons via NCS}

\begin{figure}[ht]
	\begin{center}
		\includegraphics[width=\linewidth]{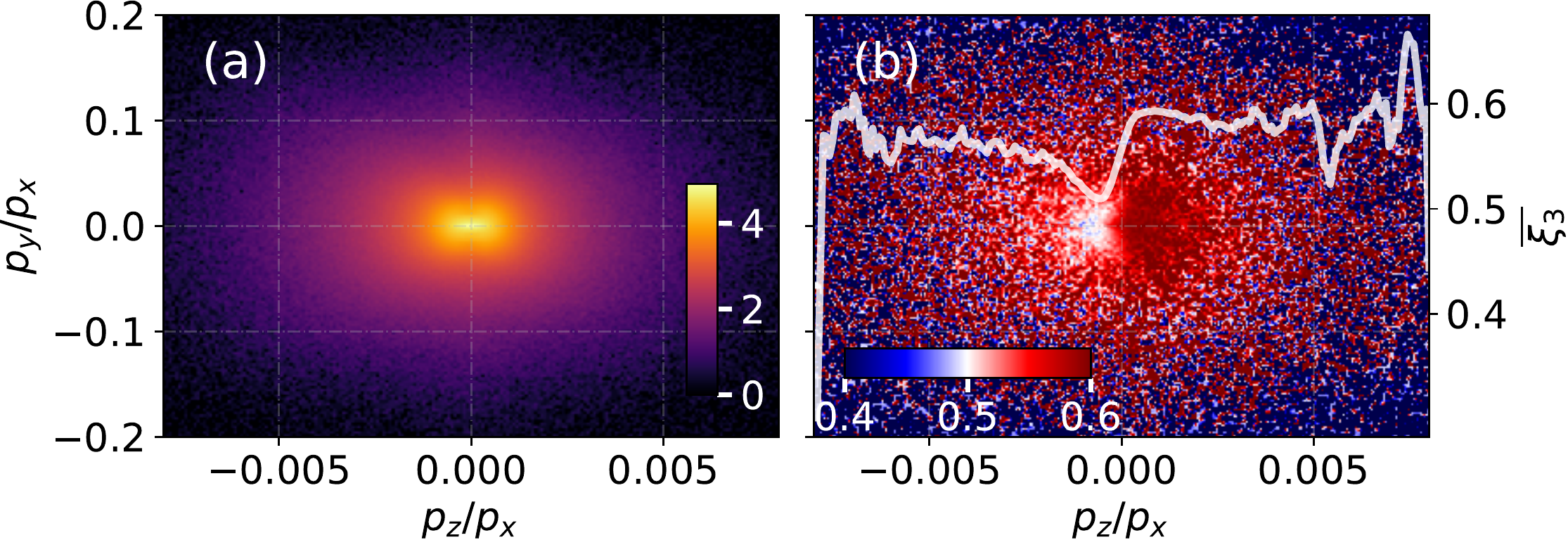}
		\caption{Generation of LP $\gamma$-photons: (a) number density $\log_{10}(d^2N/d\theta_x d \theta_y)$ (a.u.), (b) linear polarization $ \xi_3$.}
\label{lpgamma}	\end{center} 
\end{figure}

\begin{figure}[ht]
	\begin{center}
		\includegraphics[width=\linewidth]{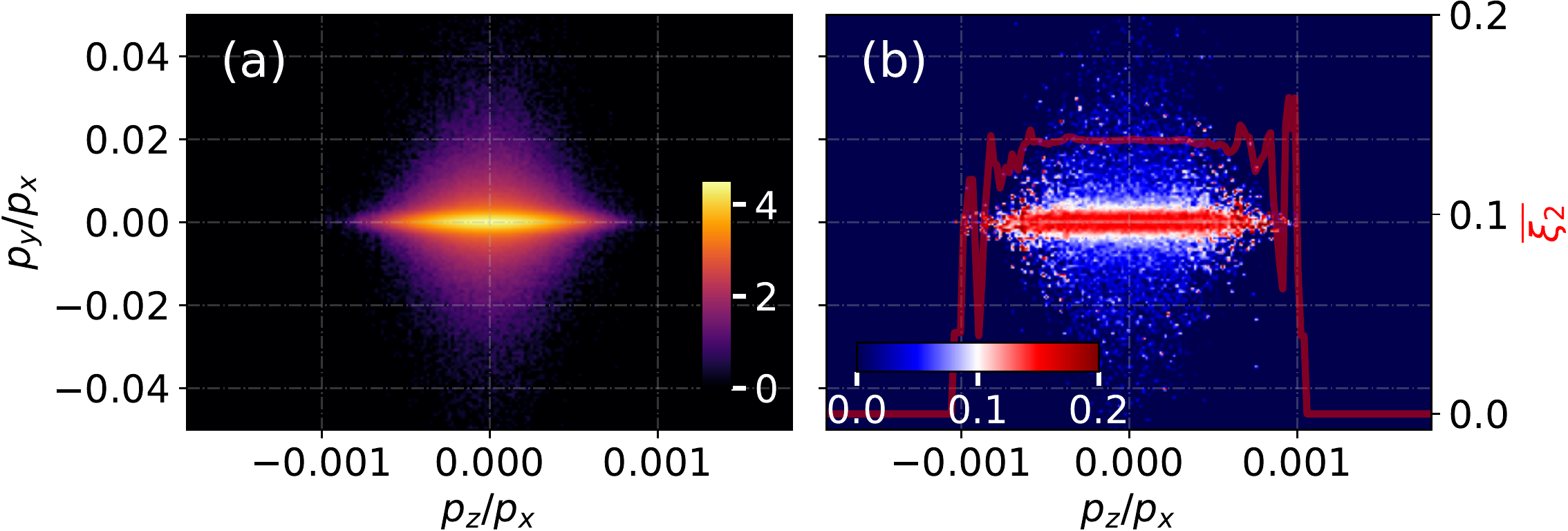}
		\caption{Generation of CP $\gamma$-photons with longitudinally polarized electrons: (a) number density $\log_{10}(d^2N/d\theta_x d \theta_y)$ (a.u.), (b) circular polarization $|\xi_2|$.} \label{cpgamma}	\end{center} 
\end{figure}

The polarization state of emitted photons can be determined in the spin/polarization-resolved NCS. 
Here, follow Ref.~\cite{Li2022}, we utilized a linearly polarized (LP) laser to collide with an unpolarized electron bunch to generate LP $\gamma$-photons. 
Additionally, we used an LP laser to collide with a longitudinally polarized electron bunch to generate circularly polarized (CP) $\gamma$-photons, which were also observed in a previous study (Ref.~\cite{Li_2020_Polarized}). The final polarization states of LP and CP $\gamma$-photons are presented in Figs.~\ref{lpgamma} and~\ref{cpgamma}, respectively.

\subsection{Laser-plasma interactions}
\begin{figure}[ht]
	\begin{center}
	\includegraphics[width=\linewidth]{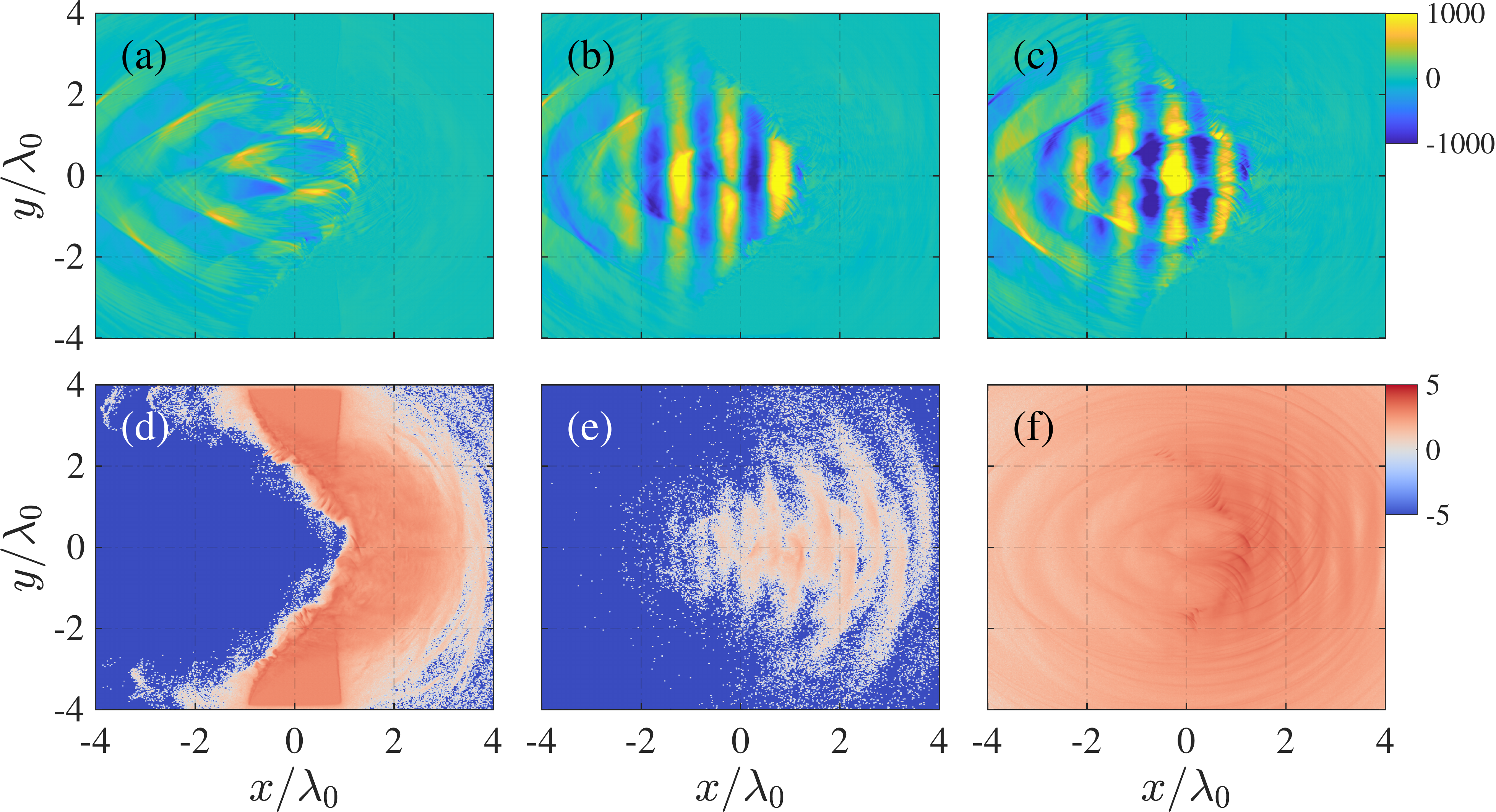}
	\caption{The laser plasma interaction via 2D simulation: spatial distribution of (a) $E_x$, (b) $E_y$ and (c) $B_z$; and the target electron distribution (d), generated NBW positron (e) and NCS $\gamma$ photon (f).}\label{laser-plasma-ebn}
	\end{center}
\end{figure}

\begin{figure}[ht]
\begin{center}
	\includegraphics[width=0.9\linewidth]{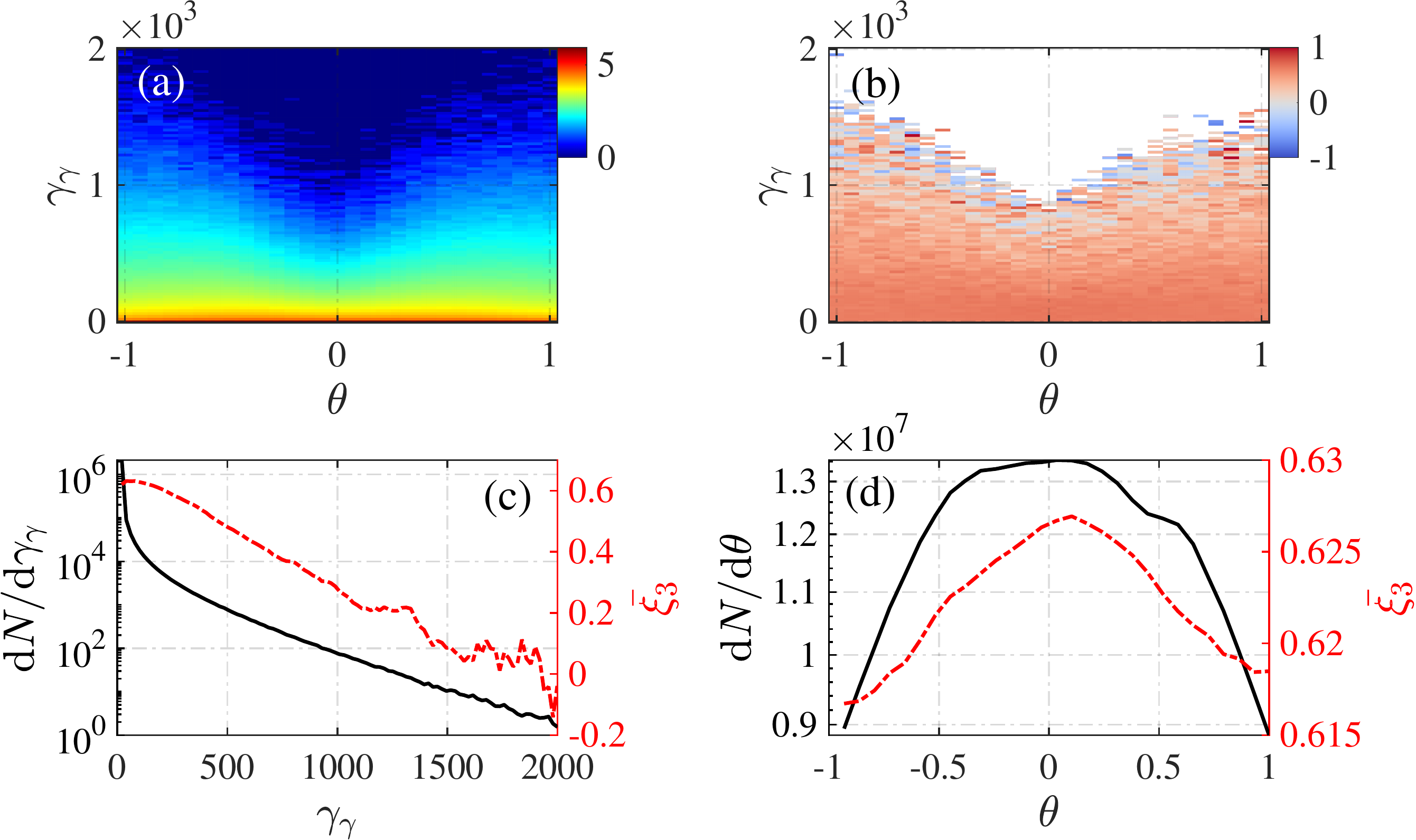}
	\caption{The laser plasma interaction generated photons: (a) number density with respect to energy and angle distribution, i.e., $\log_{10}$(d$N^2$/d$\gamma_\gamma$d$\theta$) with $\gamma_\gamma \equiv \mathcal{E}_\gamma/m_ec^2$ and $\theta\equiv p_y/p_x$, (b) energy and angle resolved linear polarization degree $\bar{\xi_3}$, (c) energy-resolved number and polarization distribution, (d) angle-resolved number and polarization distribution.}\label{laser-plasma}
\end{center}
\end{figure}

\begin{figure}[ht]
	\begin{center}
	\includegraphics[width=0.8\linewidth]{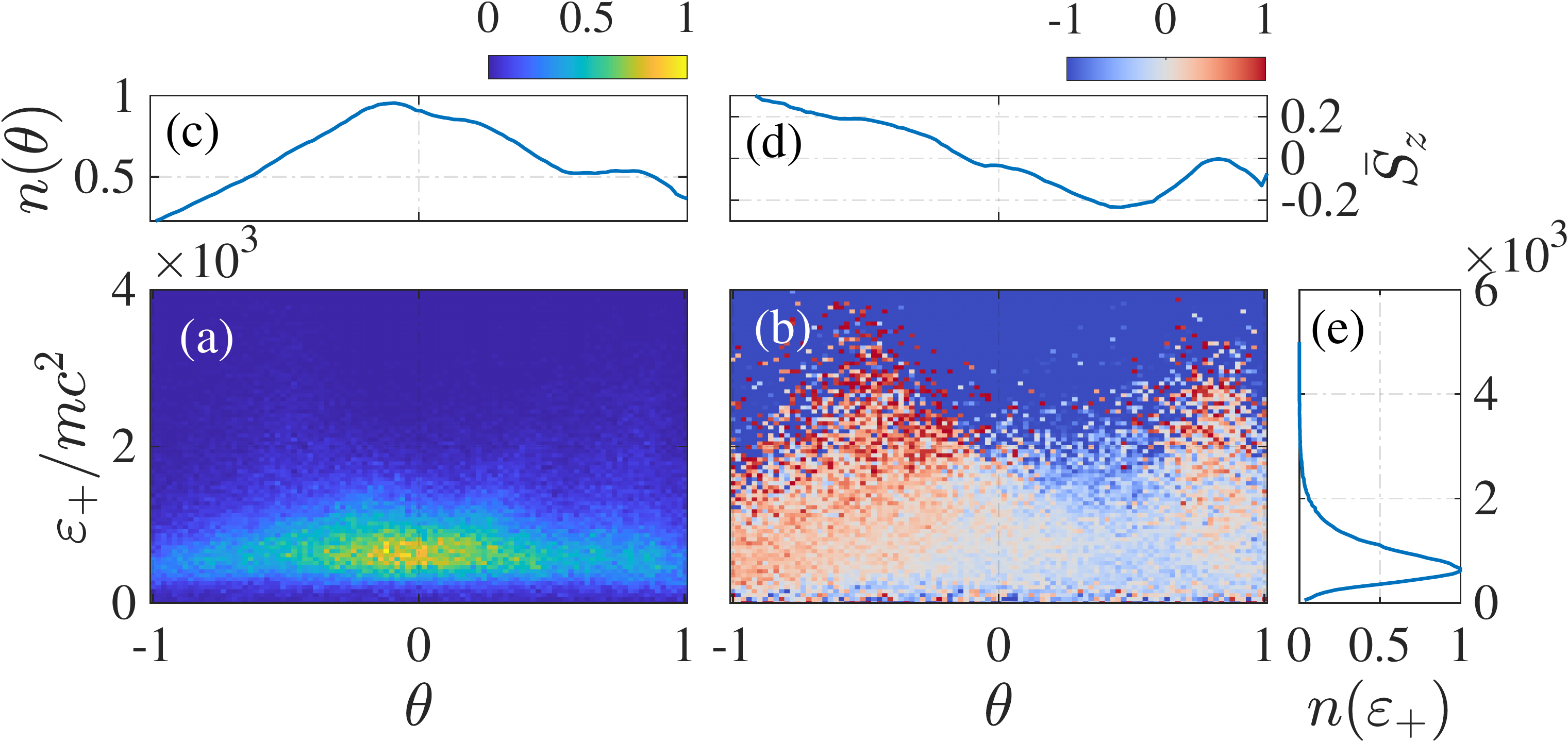}
	\caption{The laser plasma interaction generated positrons: (a) number density with respect to energy and angle distribution, i.e., d$N^2$/d$\gamma_+$d$\theta$ (in arbitrary units, arb. units) with $\gamma_+ \equiv \varepsilon_+/m_ec^2$ and $\theta\equiv \arctan(p_y/p_x)$, (b) energy and angle resolved spin component $\overline{S}_z$, (c) normalized angular distribution $n(\theta) \equiv \mathrm{d}N/\mathrm{d}\theta$, (d) angular distribution of $\overline{S}_z$, i.e., energy averaged and (e) normalized energy distribution $n(\varepsilon_+) \equiv \mathrm{d}N/\mathrm{d}\varepsilon_+$.}\label{laser-plasma-positron}
	\end{center}
\end{figure}

Finally, we present a simulation result demonstrating the interaction between an ultra-intense laser with a normalized intensity of $a_0 = 1000$ and a fully ionized $2\mathrm{\mu m}$-thick aluminum target. Note, this configuration, previously examined in Ref.~\cite{Ridgers2012} with a thickness of $1\mathrm{\mu m}$, employs a thicker target in the present study to enhance the SF-QED processes. When the laser is directed towards a solid target, the electrons experience acceleration and heating due to the laser and plasma fields. As high-energy electrons travel through the background field, they can emit $\gamma$ photons via NCS. The EM field distribution and number density of target electrons, NBW positrons, and NCS $\gamma$ photons are shown in Figure.~\ref{laser-plasma-ebn}, both of which show good consistency with Ref.~\cite{Ridgers2012}. The laser is linearly polarized along the $y$ direction, indicating that the polarization frame is mainly in the $y$-$z$ plane with two polarization bases, $\bm{e}_1 \equiv \boldsymbol{\beta} \times \boldsymbol{\dot{\beta}}$ and $\bm{e}_2 \equiv \hat{\bf n} \times \bm{e}_1$, where $\hat{\bf n}$ denotes the momentum direction of the photon.
The polarization angle-dependence observed in this study is consistent with prior literature. However, the average linear polarization degree is approximately 60\% ($\bar{\xi_3} \simeq 0.6$), as illustrated in Figs.~\ref{laser-plasma}(b) and (d). Notably, low-energy photons contribute primarily to the polarization, as demonstrated in Figs.~\ref{laser-plasma}(a) and (c). Additionally, during the subsequent NBW process, the self-generated strong magnetic field couples with the laser field dominate the positrons' SQA. As a result, the positrons' polarization is aligned with the $z$ direction, contingent on their momentum direction, as shown in Figure.~\ref{laser-plasma-positron}. These findings constitute a novel contribution to the investigation of polarization-resolved laser-plasma interactions.

\section{Outlook}
Computer simulation techniques for laser and plasma interactions are constantly evolving, not only in the accuracy of high-order or explicit/implicit algorithms but also in the complexity of new physics with more degrees of freedom. The rapid development of ultraintense techniques not only provides opportunities for experimental verification of SF-QED processes in the high-energy density regime (which serves as a micro-astrophysics lab) but also presents challenges in theoretical analysis. The introduction of Spin-QED into widely accepted PIC algorithms may address this urgent demand and pave the way for studies in laser-QED physics, laser-nuclear physics (astrophysics), and even physics beyond the standard model.

\section{Acknowlegement}
The work is supported by the National Natural Science Foundation of China (Grants No. 12275209, 12022506, and U2267204), Open Foundation of Key Laboratory of High Power Laser and Physics, Chinese Academy of Sciences (SGKF202101), the Foundation of Science and Technology on Plasma Physics Laboratory (No. JCKYS2021212008), and the Shaanxi Fundamental Science Research Project for Mathematics and Physics (Grant No. 22JSY014).

\bibliography{refs}

\end{document}